%% file: sample-sigconf.tex
\documentclass[sigconf]{acmart}
\AtBeginDocument{%
  }

\setcopyright{cc}
\setcctype{by}
\copyrightyear{2026}
\acmYear{2026}
\acmConference[WSDM 2026] {the Nineteenth ACM International Conference on Web Search and Data Mining}{ February 22--26, 2026}{Boise, ID, USA.}

\acmISBN{979-8-4007-2079-6/26/02}

\acmDOI{10.1145/XXXXXX.XXXXXX}

\usepackage{graphicx}
\usepackage{subcaption}

\usepackage{enumitem}
\usepackage{algorithm}
\usepackage{algorithmicx}
\usepackage{algpseudocode}
\usepackage{url}
\usepackage{booktabs}
\floatname{algorithm}{Algorithm} 

\usepackage{color}
\definecolor{brown}{RGB}{139,64,0}

\begin{document}

\title{Continuous-time Discrete-space Diffusion Model for Recommendation}

\author{Chengyi Liu}
\orcid{0000-0002-3463-3055}
\email{chengyi.liu@connect.polyu.hk}
\affiliation{%
  \institution{The Hong Kong Polytechnic University}
  \city{Hong Kong SAR}
  \country{China}
}

\author{Xiao Chen}
\affiliation{%
  \institution{The Hong Kong Polytechnic University}
  \city{Hong Kong SAR}
  \country{China}}
\email{shawn.chen@connect.polyu.hk}

\author{Shijie Wang}
\affiliation{%
  \institution{The Hong Kong Polytechnic University}
  \city{Hong Kong SAR}
  \country{China}}
\email{shijie.wang@connect.polyu.hk}

\author{Wenqi Fan}
\affiliation{%
  \institution{The Hong Kong Polytechnic University}
  \city{Hong Kong SAR}
  \country{China}}
\email{wenqi.fan@polyu.edu.hk}

\author{Qing Li}
\affiliation{%
  \institution{The Hong Kong Polytechnic University}
  \city{Hong Kong SAR}
  \country{China}}
\email{csqli@comp.polyu.edu.hk}
\renewcommand{\shortauthors}{Chengyi Liu, Xiao Chen, Shijie Wang, Wenqi Fan, and Qing Li}

\input{sections/abstract}






\maketitle

\input{sections/Introduction}

\input{sections/Method}

\input{sections/Experiment}
\input{sections/RelatedWork}
\input{sections/Conclusion}









\bibliographystyle{ACM-Reference-Format}
\bibliography{sample-base}



\end{document}

%% file: sections/abstract.tex
\begin{abstract}
In the era of information explosion, Recommender Systems (RS) are essential for alleviating information overload and providing personalized user experiences. Recent advances in diffusion-based generative recommenders have shown promise in capturing the dynamic nature of user preferences. These approaches explore a broader range of user interests by progressively perturbing the distribution of user-item interactions and recovering potential preferences from noise, enabling nuanced behavioral understanding. However, existing diffusion-based approaches predominantly operate in continuous space through encoded graph-based historical interactions, which may compromise potential information loss and suffer from computational inefficiency. As such, we propose CDRec, a novel Continuous-time Discrete-space Diffusion Recommendation framework, which models user behavior patterns through discrete diffusion on historical interactions over continuous time. The discrete diffusion algorithm operates via discrete element operations (e.g., masking) while incorporating domain knowledge through transition matrices, producing more meaningful diffusion trajectories. Furthermore, the continuous-time formulation enables flexible adaptive sampling. To better adapt discrete diffusion models to recommendations, CDRec introduces: (1) a novel popularity-aware noise schedule that generates semantically meaningful diffusion trajectories, and (2) an efficient training framework combining consistency parameterization for fast sampling and a contrastive learning objective guided by multi-hop collaborative signals for personalized recommendation. Extensive experiments on real-world datasets demonstrate CDRec's superior performance in both recommendation accuracy and computational efficiency.
Our codes are available at: ~\url{https://github.com/ChengyiLIU-cs/CDRec}

\end{abstract}

%% file: sections/Introduction.tex
\section{Introduction}
\label{Introduction}

With the rapid growth of digital content, Recommender Systems (RS) have become essential to alleviate information overload and enhance user experience by providing personalised content across diverse scenarios, such as e-commerce and social media ~\cite{wang2019neural, fan2019graph}.
Technically, as one of the most representative RS techniques,  Collaborative Filtering (CF) methods aim to infer user preferences from user-item interaction behaviours in the history~\cite{zhao2024recommender,qu2024ssd4rec}. 
Graph Neural Network (GNN)-based approaches in CF have achieved notable success by leveraging their capability to model graph-structured data~\cite{he2020lightgcn, fan2022graph, wu2021self, chen2023fairly}. 
These methods refine user and item representations on historical interaction graphs, capturing high-order collaborative signals within a discriminative recommendation paradigm ~\cite{yang2023generate}. 
Recent studies have focused on generative models that learn user interaction distributions to directly generate preferred items, effectively modeling the temporal dynamics of user preferences ~\cite{wang2023generative, ning2024cheatagent, qu2025generative, wang-etal-2025-knowledge-graph, qu2025tokenrec, wang2025graph}. 
In particular, diffusion-based generative recommender models frame the recommendation task as simulating real-world interaction processes from noise, offering a promising solution ~\cite{wang2023diffusion, choi2023blurring, liu2024score, he2025graph}.

\begin{figure}[t!]
\centering
\includegraphics[width=0.48\textwidth]{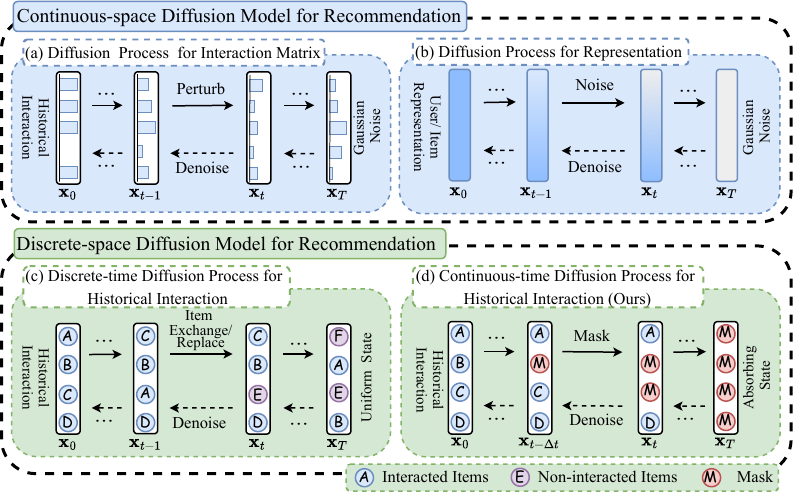}%
\caption{Comparison of diffusion-based recommender systems across different state-spaces. 
Continuous-space diffusion algorithms apply isotropic Gaussian noise to either the collaborative graph’s adjacency matrix or the encoded representations, often compromising personalized information in the embedding space. 
In contrast, discrete-space diffusion models directly operate through item-level operations toward a uniform distribution. 
Inspired by this formulation, our proposed CDRec method perturbs historical interactions via masking operations in continuous time, enabling state transitions to occur at arbitrary time steps.
}
\label{fig:RecDiff} 
\vskip -0.15in
\end{figure}

Diffusion-based recommendation methods exhibit notable potential due to their ability to model complex data distributions and generate samples with broad coverage ~\cite{ho2020denoising, song2021scorebased}. 
Grounded in a solid theoretical foundation, the typical diffusion algorithm first perturbs the original data distribution into a known prior and consequently learns a parameterized reverse process to iteratively construct samples from noise. 
This diffusion mechanism aligns well with the objective of RS, which is to infer the distribution of potential user-item interactions from inherently incomplete and noisy historical data. 
To be specific, most existing methods directly apply continuous-space diffusion models on discrete graph-structured data, which can be broadly categorized into two types: \emph{applying diffusion algorithms either on the user-item interaction matrix or on user/item representations.} 
For instance, DiffRec ~\cite{wang2023diffusion} models user preferences by directly perturbing binary-encoded historical interactions and reconstructing the collaborative graph’s adjacency matrix. 
In contrast, methods like DreamRec ~\cite{yang2023generate} initiate from Gaussian noise and progressively generate user and item representations in continuous space. 
By decomposing preference modeling into a multi-step denoising process, diffusion models enhance the capability to capture complex user-item interactions ~\cite{zhao2024denoising}.

Despite their demonstrated success, conventional diffusion models operating in continuous state-space face notable limitations when directly applied to discrete CF data. 
These algorithms typically introduce isotropic Gaussian noise to perturb encoded graph representations, resulting in undesirable information loss. 
As illustrated in Figure~\ref {fig:RecDiff} (a), approaches that perturb the interaction matrix apply uniform noise across all items, ignoring item dependencies and disrupting the collaborative graph's topological structure~\cite{zhu2024graph}.
In Figure~\ref{fig:RecDiff} (b), methods that corrupt user/item representations with Gaussian noise risk degrading semantic consistency, leading to uninformative diffusion trajectories that hinder effective training and sampling~\cite{xie2024breaking}.

To overcome these limitations, we investigate diffusion models in discrete state-spaces\footnote{For brevity, “discrete-space diffusion” and “discrete diffusion” are interchangeable terms for diffusion models in discrete state-spaces.}, which have demonstrated significant success in discrete text generation yet remain unexplored in the context of recommender systems.
These algorithms operate via state transitions, such as element swapping or replacement, progressively transforming discrete data toward a stationary distribution (Figure \ref{fig:RecDiff} (c)).
More importantly, the discrete forward process facilitates direct integration of domain knowledge through transition matrices, yielding more informative diffusion trajectories for recommendation learning~\cite{austin2021structured}.
This formulation fits the recommendation setting, as it directly models interaction sequences, rather than treating them merely as conditional information ~\cite{xie2024breaking}.
Despite its great potential, directly applying discrete diffusion to recommendation tasks is highly challenging. 
First, simply perturbing user interactions toward a uniform distribution remains semantically meaningless for recommendation tasks, potentially degrading user preference information.
Therefore, the discrete diffusion algorithm requires careful design that incorporates domain-specific knowledge (e.g., item popularity) to preserve personalized signals. 
Moreover, current theoretical foundations for parameterizing the reverse process remain developing, necessitating tailored adaptation for recommendation tasks to ensure efficient training and generation ~\cite{lou2024discrete}. 
Additionally, establishing effective collaborative signal guidance is essential to learn interaction distributions for personalized recommendations.

In this work, we propose a novel \underline{C}ontinuous-time \underline{D}iscrete-space Diffusion \underline{Rec}ommendation framework (CDRec) to address the above challenges. The proposed CDRec employs an absorbing state to perturb historical interactions via a \textbf{masking operation}, and learns a parameterized reverse process for personalized recommendation, as shown in Figure \ref{fig:RecDiff} (d). 
Compared to the discrete-time algorithm, the continuous-time framework, formulated through a stochastic differential equation (SDE), enables flexible state transitions at arbitrary time intervals. 
This enhances modeling of user behavior patterns while generalizing beyond conventional ancestral sampling strategies ~\cite{campbell2022continuous}.
Specifically, the proposed CDRec proposes a novel popularity-aware noise schedule, where items with higher interaction frequencies are assigned lower absorption probabilities during the forward diffusion process. 
This schedule encourages the diffusion model to simulate real-world interaction processes during reverse diffusion, which starts interaction from popular items, and thereby facilitates more accurate preference modeling.
Instead of estimating the reverse transition rate, CDRec parameterizes the reverse diffusion process using a consistency function that models user behavior patterns over masked historical interactions. 
This formulation enables a balance between sampling quality and efficiency by supporting both single-step and multi-step generation.

Our main contributions are summarized as follows:
\begin{itemize} [leftmargin=*]
    \item We propose a novel recommendation framework (\textbf{CDRec}) that incorporates discrete diffusion processes to support efficient recommendation in continuous-time steps.
    
    \item To enhance training and sampling efficiency in the diffusion process, we introduce a novel popularity-aware noise schedule that generates informative diffusion trajectories by simulating real-world interaction dynamics. 
    
    \item We design a consistency function to parameterize the reverse diffusion process, combined with a contrastive learning objective that integrates collaborative signals to guide the diffusion process for recommendation. 
    
    \item Extensive experiments on three real-world datasets demonstrating our proposed CDRec’s superior performance over state-of-the-art baselines, with ablation studies verifying each component’s contribution. 
    
\end{itemize}

%% file: sections/Method.tex
\section{The Proposed Method}
\label{Method}

In this section, we first introduce the key notations and definitions used throughout the paper. We then present an overview of the proposed framework, CDRec, followed by a detailed description of its components.

\subsection{Notations and Definitions}

In recommender systems, let $\mathcal{U}$ denote the set of $n$ users and $\mathcal{V}$ indicate the set of $m$ items. For each user $u \in \mathcal{U}$, the historical interactions are represented as $\mathbf{x}_u = [\mathbf{v}_1, \mathbf{v}_2, \dots, \mathbf{v}_l]$, where $l$ is the predefined sequence length and each $\mathbf{v}_i \in \mathcal{V}$ is represented in one-hot form. 
To model collaborative signals, we represent users and items through learnable embeddings, where $\mathbf{P} = [\mathbf{p}_1, \dots, \mathbf{p}_n]^T \in \mathbb{R}^{n \times d}$ denotes the user embedding matrix and $\mathbf{Q} = [\mathbf{q}_1, \dots, \mathbf{q}_m]^T \in \mathbb{R}^{m \times d}$ represents the item embedding matrix. These embeddings can be initialized using any arbitrary pre-trained GNN-based recommendation model.

\begin{figure*}[thp] 
\centering
\includegraphics[width=0.99\textwidth]{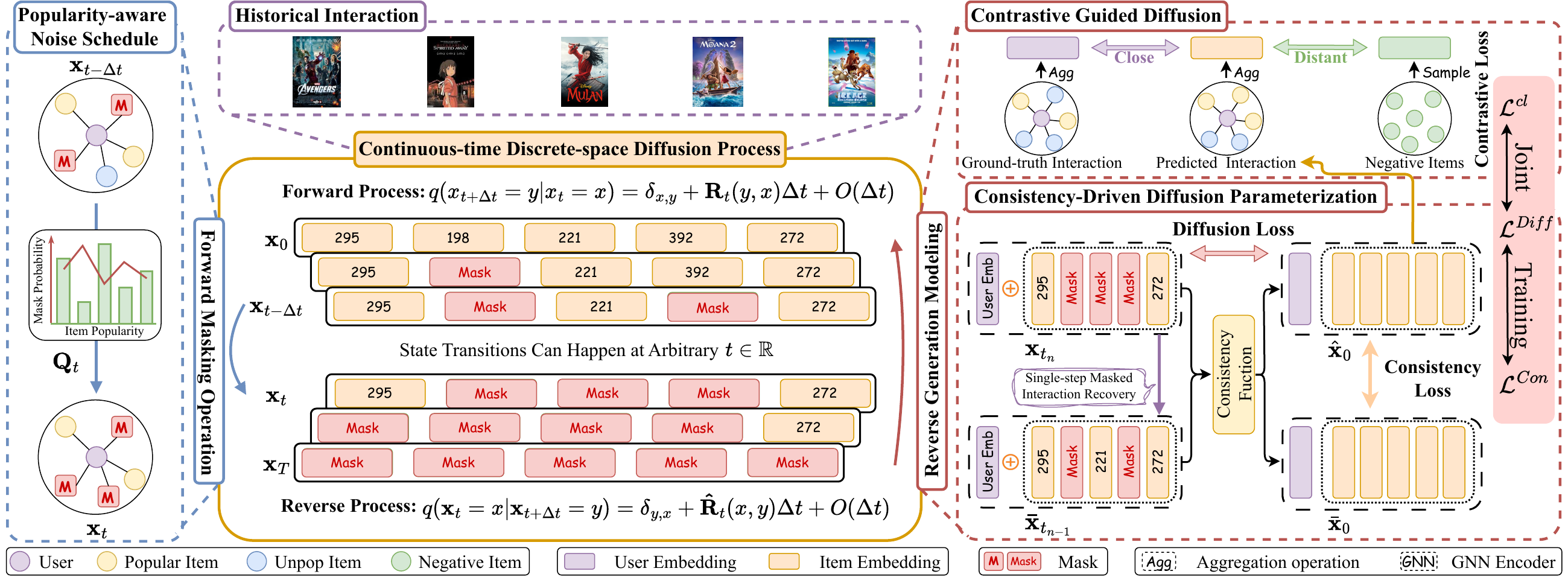}%
\caption{
The proposed CDRec framework comprises three main modules: popularity-aware noise schedule, consistency-driven diffusion parameterization, and contrastive-guided diffusion. The left panel illustrates the forward diffusion process, where state transitions are controlled by item popularity to simulate interaction dynamics. The right panel presents the reverse generation modeling process, which leverages a consistency function for efficient parameterization and applies contrastive learning to guide personalized recommendations with the structural collaborative signal.  
}
\label{fig:diffrec} 
\end{figure*}

\subsection{An Overview of the Proposed Framework}
We propose a continuous-time framework that directly models the distribution of historical interactions using the discrete-space diffusion algorithm. 
Specifically, CDRec initiates the forward diffusion process with a popularity-aware noise schedule that perturbs interactions via masking operations, progressively transitioning toward an absorbing state to simulate real-world interaction dynamics. In the reverse generation phase, the framework parameterizes the reverse process with a consistency function, enabling both efficient one-step generation and high-quality iterative sampling. To enhance the personalized recommendation, we incorporate contrastive learning with structural collaborative signals to guide the sampling process.
Recommendations are generated by propagating synthesized items as user representations and selecting top-$k$ items based on user–item embedding similarity. This approach fully exploits the generative capacity of diffusion models over interaction sequences, rather than treating them solely as conditional input.
The overall framework is demonstrated in Figure~\ref{fig:diffrec}.

\subsection{Popularity-Aware Discrete Diffusion Algorithm}

To mitigate information degradation in existing diffusion recommender models, we introduce a novel popularity-aware noise schedule to control the forward diffusion process, enabling the model to learn item distributions across popularity levels at different diffusion phases. This approach integrates prior knowledge to construct informative diffusion trajectories, thereby enhancing both training and sampling.

\subsubsection{Forward Diffusion Process.}
The forward process constructs a Continuous-Time Markov Chain to transform the original data distribution $p_0$ into a stationary distribution $p_{ref}$. 
Formally, we model the user $u$'s historical interactions $\mathbf{x}_u \sim p_0$ as the initial state, with an absorbing state serving as the prior distribution ~\cite{austin2021structured}. During forward diffusion, each item $\mathbf{v}_i$ is denoted as $\mathbf{x}_0 \in \mathbb{R}^{1\times|\mathcal{V}|}$ under the SDE framework, and undergoes progressive replacement by a [MASK] token.
The absorbing state distribution preserves the real user-interacted data throughout the diffusion trajectory, thereby providing more meaningful learning signals than corrupted embeddings ~\cite{xie2024breaking}.
The transition between time points $t$ and $t+\Delta{t}$ can be described by the following SDE:
\begin{equation}
    q(\mathbf{x}_{t+\Delta{t}} = y|\mathbf{x}_{t} = x)= \delta_{x,y}+\mathbf{R}_{t}(y,x)\Delta{t}+O(\Delta{t}),
    \label{eq:forward}
\end{equation}
where $x$ and $y$ represents the item state at time $t$ and $t+\Delta{t}$ respectively, $\delta$ refers to the Kronecker delta, $\mathbf{R}_t \in \mathbb{R}^{|\mathcal{V}|\times|\mathcal{V}|}$ denotes the transition rate matrix, and $O(\Delta{t})$ refers to the terms whose magnitude becomes negligible relative to $\Delta {t}$ in the limit as $\Delta {t} \rightarrow 0$. 
This absorbing behavior is determined by the transition rate matrix $\mathbf{R}_{t}$ assigned to each item, where higher transition rates correspond to shorter absorption times.

According to the law of total probability, the marginal distribution $p_t(\mathbf{x})$ at time $t$ is given by
~\cite{oksendal2003stochastic}:
\begin{equation}
    p_t(\mathbf{x}) = \int p_0(\mathbf{x}_0)q_{t|0}(\mathbf{x}_t | \mathbf{x}_0)d\mathbf{x}_0.
\end{equation}
To derive an efficient transition kernel $q_{t|0}(\mathbf{x}_t| \mathbf{x}_0)$, CDRec adopts the standard formulation $\mathbf{R}_{t} = \beta(t) \mathbf{R}$, where $\beta(t)$ refers to the noise schedule, and $\mathbf{R}$ denotes the manually defined fixed base rate. Consequently, the analytical solution is derived as follows:
\begin{equation}
\begin{aligned}
     q_{t|0}(\mathbf{x}_t = j| \mathbf{x}_0 = i) &\ = \left(\mathbf{S}  \text{exp}(\Lambda \overline{\beta}(t) ) \mathbf{S}^{-1}   \right)_{ij}, \\
     \overline{\beta}(t) &\ = \int_{0}^{t} \beta_t dt,
     \label{eq:trasition_kernal}
\end{aligned}
\end{equation}
where $\mathbf{R} = \mathbf{S} \Lambda \mathbf{S}^{-1}$ is the eigendecomposition of the base rate matrix $\mathbf{R}$, with $\mathbf{S}$ containing the eigenvectors of $\mathbf{R}$ and $\Lambda$  being the diagonal matrix of eigenvalues. The $\text{exp}(\cdot)$ denotes the element-wise exponential function, and $i$, $j$ index distinct states.

\subsubsection{Popularity-aware Noise Schedule.}
Existing approaches predominantly adopt linear or cosine schedules for $\beta(t)$, which are suboptimal for recommendation tasks. These conventional schedules typically assume uniform item contributions to preference modeling, thereby overlooking important inter-item correlations~\cite{sun2022score}.
To address this limitation and construct more informative diffusion trajectories, we introduce a novel sequence-level noise schedule that simulates real-world interaction patterns by prioritizing popular items \cite{huang2022exploring}.
Since popular items typically reflect current trends and better characterize user behavior, they provide particularly meaningful information for interaction modeling ~\cite{zhang2021causal}.
Consequently, our method retains high-frequency items for extended durations during diffusion, thereby enabling more precise modeling of user preferences.

\begin{figure}[bhp]
\centering
\includegraphics[width=0.35\textwidth]{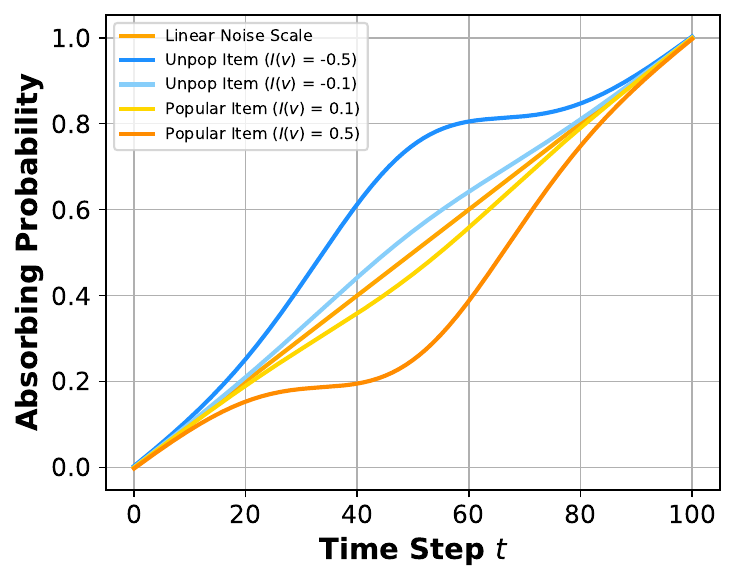}%
\caption{Absorbing probability for items with varying popularity deviation ($\omega=0.5$).
}
\label{fig:absorb} 
\vskip -0.13in
\end{figure}

We formulate a popularity deviation metric $I(\cdot)$ that measures the relative popularity of an item $\mathbf{v}$ within the user's interaction sequence $\mathbf{x}_u$:
\begin{equation}
    I(\mathbf{v}) = \text{pop}(\mathbf{v}) - \frac{\sum_{0}^{l}{\text{pop}(\mathbf{v}_i)}}{l},
\end{equation}
where $\mathbf{v}_i$ denotes $i$-th interacted item in the sequence $\mathbf{x}_u$, and $\text{pop}(\mathbf{v}) \in [0,1]$ denotes the frequency of item $\mathbf{v}$, normalized within the interaction sequence.
Our design captures item correlations with respect to popularity by assigning positive deviation values to items with above-average interaction frequencies and negative values otherwise.
Then the absorbing probability of each item $\mathbf{v}_i$ is defined as :
\begin{equation}
    \overline{\beta}(t)_{\mathbf{v}_i} = \frac{t}{T} - \omega \  \text{exp}\left(- \frac{t-\frac{T}{2}}{2\sigma^{2}} \right)  I(\mathbf{v}_i),
    \label{eq:noise_schedule}
\end{equation}
where $\omega$ and $\sigma$ are hyperparameters to scale the noise schedule at time step $t$.
The popularity deviation effect is adaptively scaled according to a normal distribution, as illustrated in Figure ~\ref{fig:absorb}. 
Compared to linear schedules, this yields lower absorbing probabilities for relatively popular items and higher probabilities for unpopular ones.
Our design guarantees that higher-frequency items ($I(\mathbf{v}_i) > I(\mathbf{v}_j)$) experience lower corruption rates ($\overline{\beta}(t)_{\mathbf{v}_i} < \overline{\beta}(t)_{\mathbf{v}_j}$) during forward diffusion.
This creates an easy-to-hard reverse process, as popular items inherently exhibit more stable patterns~\cite{zhang2021causal}.
This staged masking strategy adapts to popularity variations across diffusion stages, allowing the model to progressively learn items of varying popularity at different phases, thereby effectively capturing the interaction distribution.

Moreover, our popularity-aware noise schedule satisfies the standard transition rate requirements ~\cite{campbell2022continuous}: (1) guaranteed convergence of the data distribution to the terminal distribution, and (2) efficient computation of the transition kernel $q_{t|0}(\mathbf{x}_t| \mathbf{x}_0)$ through Equations ~\ref{eq:trasition_kernal} and ~\ref{eq:noise_schedule}, with the structured matrix $\mathbf{R}$ defined as:

\begin{equation}
\mathbf{R} =
\begin{bmatrix}
-1 & 0 & \cdots & 0 & 0 \\
0 & -1 & \cdots & 0 & 0 \\
\vdots & \vdots & \ddots & \vdots & \vdots \\
0 & 0 & \cdots & -1 & 0 \\
1 & 1 & \cdots & 1 & 0 \\
\label{eq:matrix}
\end{bmatrix}.
\end{equation}

\subsubsection{Reverse Diffusion Process}
The reverse SDE learns to generate personalized item interactions through iterative denoising from the absorbing state, effectively modeling the underlying dynamics of real user preference evolution.
The reverse-time SDE can be formulated as:
\begin{equation}
    q(\mathbf{x}_{t} = x|\mathbf{x}_{t+\Delta{t}} = y)= \delta_{y,x}+\mathbf{\hat{R}}_{t}(x,y)\Delta{t}+O(\Delta{t}),
    \label{eq:reverse}
\end{equation}
where $\mathbf{\hat{R}}_{t} \in \mathbb{R}^{|\mathcal{V}|\times|\mathcal{V}|}$ represents the reverse transition rate matrix. 
Considering both the forward and reverse diffusion process are Markovian, the $\mathbf{R}_{t}$ and $\mathbf{\hat{R}}_{t}$ satisfy the following correlation:
\begin{equation}
    \mathbf{\hat{R}}_{t}(x, y) = \frac{q_{t}(y)}{q_{t}(x)} \mathbf{R}_{t}(y, x),
    \label{eq:score}
\end{equation}
where $q_{t}(y)$ and ${q_{t}(x)}$ denote the marginal probabilities of state $x$ and $y$ at time $t$ , respectively. The ratio $q_{t}(y) / {q_{t}(x)}$ serves as a concrete score, analogous to the score function $\nabla_{\mathbf{x}}  \log p_t$ in continuous-state diffusion algorithms ~\cite{sun2022score}. However, this concrete score is generally intractable and requires accurate parameterization to enable the reverse diffusion generation process.

\subsection{Consistency-Driven Diffusion Parameterization}

Current parameterization methods for discrete-space diffusion remain developing and can be broadly categorized into two classes: modeling the reverse density $p_{0|t}$ and modeling the score $q_{t}(y) / {q_{t}(x)}$, both of which face inherent limitations in the recommendation context ~\cite{lou2024discrete}. Direct estimation of $p_{0|t}$ often yields suboptimal performance due to discrete distributions ~\cite{campbell2022continuous}, while learning the score typically requires substantial computational resources and struggles to ensure positive transition rates during the training process, potentially introducing undesired bias ~\cite{meng2022concrete, sun2022score}.

To address these limitations, we explore an alternative consistency function for parameterizing discrete diffusion processes, inspired by consistency models ~\cite{song2023consistency}. Our parameterization enables efficient single-step generation while supporting iterative refinement to balance sample quality and computational cost.

\subsubsection{Consistency Function} 
The consistency function learns the diffusive trajectory of the masked historical interaction $\{\mathbf{x}_{t}\}_{t\in [\varepsilon, T]}$, bypassing explicit estimation of the reverse transition rate $\mathbf{\hat{R}}_{t}$. It directly outputs predicted user-item interactions by reconstructing the original preference distribution conditioned on the collaborative signal.
Following the standard formulation ~\cite{geng2024consistency}, we define the consistency function as follows:
\begin{equation}
    f: (\mathbf{x}_t,t)\mapsto \mathbf{x}_{\varepsilon},
\end{equation}
where $\varepsilon$ denotes a fixed small positive time point, prior to which no state transitions occur. 

This formulation naturally aligns with recommendation objectives by recovering latent user-item interactions.
However, the well-established skip-connection architectures for consistency functions are fundamentally incompatible with discrete-space diffusion, as superposition operations on discrete states $\{\mathbf{x}_{t}\}_{t\in [\varepsilon, T]}$ lack semantic validity and implementation feasibility ~\cite{song2023consistency}. 
Thus, we parameterize the consistency model in a manner that aligns with the discrete nature of CDRec as
\begin{align}
f_{\mathbf{\theta}}(\mathbf{x},t)
\begin{cases}
\ \mathbf{x}, & t = \varepsilon\\
\ F_{\mathbf{\theta}}(\mathbf{x},t), & t \in (\varepsilon, T] 
\end{cases}\label{eq:SDE_ins}
\end{align}
where $F_{\mathbf{\theta}}$ is a free-form deep neural network. 

\subsubsection{Consistency Training Strategy}
The core of training the consistency function lies in minimizing the difference between outputs from data pairs at adjacent time points, $(\mathbf{\bar{x}}_{t-\Delta{t}}, \mathbf{x}_{t})$, where $\mathbf{\bar{x}}_{t_{n-1}}$ is estimated from $\mathbf{x}_{t_n}$, enabling $f_\mathbf{\theta}$ to estimate user preferences from samples at arbitrary time steps ~\cite{geng2024consistency}. In the context of discrete-space diffusion, this strategy is inefficient, as state transitions may not occur within the infinitesimal interval $\Delta{t}$. To address this, we denote the current time step $t \sim \mathbf{\mathcal{U}}(\varepsilon, T)$ as $t_n$, and we define $t_{n-1} = t_n - \nabla {t}$, where $\nabla {t}$ represents the fixed time interval. The training objective $\mathcal{L}^{Con}$, tailored for discrete-space diffusion, is formulated as:
\begin{equation}
    \mathcal{L}^{Con} = \min_{\boldsymbol{\theta}} \mathbb{E}_{\mathbf{x}_0,{t_n}}[\gamma(t_n) d(f_{\mathbf{\theta}}(\mathbf{{x}}_{t_n},t_n),f_{\mathbf{\theta}^{-}}(\mathbf{\bar{x}}_{t_{n-1}},{t_{n-1}})],
    \label{eq:consistency_loss}
\end{equation}
where $\gamma(\cdot)$ is a time-dependent weighing function, $d(\cdot,\cdot)$ denotes a distance metric between the outputs, and $\mathbf{\theta}^{-}$ represents the exponential moving average of past values of $\mathbf{\theta}$, updated via $\mathbf{\theta}^{-} \leftarrow {stopgrad}(\mu\mathbf{\theta}^{-} + (1-\mu)\mathbf{\theta})$ to stabilize the training process.
The primary difficulty in applying this training objective lies in the efficient estimation of $\mathbf{\bar{x}}_{t_{n-1}}$. 
In the context of recommendation, no pre-trained model exists to approximate the reverse transition kernel $p(\mathbf{\bar{x}}_{t_{n-1}}|\mathbf{x}_{t_n})$, and training a dedicated model would be computationally expensive. Moreover, unlike continuous-state SDEs, discrete-space diffusion algorithms lack an unbiased estimator for the score $q_{t}(y) / {q_{t}(x)}$ ~\cite{meng2022concrete, lou2024discrete}.

Since the discrete process operates through scheduled item masking, the transition kernel $p(\mathbf{\bar{x}}_{t_{n-1}}|\mathbf{x}_{t_n})$ can be simplified via masked element recovery algorithms, bypassing complex mathematical derivations. To enable efficient training of the consistency function for the discrete-space diffusion, we propose two methods to construct the data pair $(\mathbf{\bar{x}}_{t_{n-1}}, \mathbf{x}_{t_n})$ by transit masked items to original state from sampled $\mathbf{x}_{t_n}$. 
The first method, termed one-step recovery, aims to sample $\mathbf{\bar{x}}_{t_{n-1}}$ by recovering only one masked item for simplicity. Given a sample $\mathbf{x}_{t_n}$ drawn from the transition density $q_{t|0}(\mathbf{x}_{t_n}| \mathbf{x}_0)$, we obtain the masking probabilities for each position as  $\left[\overline{\beta}(t_n)_{\mathbf{v}_1}, \overline{\beta}(t_n)_{\mathbf{v}_2}, \dots, \overline{\beta}(t_n)_{\mathbf{v}_l}\right]$. To construct $\mathbf{\bar{x}}_{t_{n-1}}$, we simply recover the masked item with the lowest masking probability. This method is simple and efficient, but requires careful design of the time interval $\nabla t$ to mitigate exposure bias ~\cite{ning2023elucidating}, which refers to the behavioral gap between training and sampling. Consequently, recovering only one item may not accurately reflect the data distribution at time step $t_{n-1}$.
The second method adopts a pseudo-Euler approach to adaptively estimate the unmasking probabilities of items over the time interval $\nabla t$ ~\cite{sun2022score}. The reverse transition probability for each item is defined as:

\begin{align}
p_{t_{n-1|t_{n}}}(\mathbf{x}^{\mathbf{v}_i}_{t_{n-1}} = x | \mathbf{x}^{\mathbf{v}_i}_{t_n} = y)
\begin{cases}
\ 1-(B_{t_n}-\frac{\nabla t}{T}B_{t_n}), & \text{if } x \neq [M], y = [M]\\
\ B_{t_n}-\frac{\nabla t}{T}B_{t_n}, & \text{if } x = y = [M]  \\
\ 1 & \text{if } x = y \neq [M],
\end{cases}\label{eq:SDE_ins}
\end{align}
where $B_{t_n} = \overline{\beta}(t_n)_{\mathbf{v}_i}$ for brevity, and $[M]$ represents the absorbing state.  This method focuses on recovering the masked items, with their probabilities scaled according to the time interval and the original masking probability, aligning well with the forward diffusion process.
Given the data pair $(\mathbf{\bar{x}}_{t-\Delta{t}}, \mathbf{x}_{t})$, we can efficiently train the consistency function $f_{\theta}$. This enables single-step generation while maintaining iterative sampling capability, as detailed in Algorithm \ref{alg:sampling}.

The $\mathcal{L}^{Con}$ enforces identical predictions under data perturbations to model the interaction distribution in a step-wise manner, but it may over-emphasize popular items, as they tend to persist longer in the diffusive trajectory. To mitigate this bias, we empirically introduce a diffusion loss $\mathcal{L}^{Diff}$ as an auxiliary denoising objective, directly modeling historical interactions through the following formulation:
\begin{equation}
    \mathcal{L}^{Diff} = -\sum_{i=1}^{l}\left(\mathbf{x}_u^{i} \log p_{\mathbf{\theta}}^{i}(\mathbf{{x}}_{0}|\mathbf{{x}}_{t}) \right),
    \label{eq:diffusion_loss}
\end{equation}
where $p_{\mathbf{\theta}}^{i}$ represents the predicted categorical probability for $i$-th element generated by the consistency model $F_{\mathbf{\theta}}$, $\mathbf{x}_u^{i}$ denotes the ground-truth value of $i$-th element. The experiments demonstrate that this loss function stabilizes the training process while enhancing the model’s capacity, consistent with previous studies~\cite{ho2020denoising, austin2021structured}.

\subsubsection{Practical Implementation}
In practice, the consistency model $F_{\mathbf{\theta}}$ is instantiated as a Transformer encoder. 
For a user $u$ with perturbed interactions $\mathbf{x}_t$ at time step $t$, the encoded interaction sequence is prefixed with a collaborative user embedding to preserve personalized information: $\langle \mathbf{p}_u, \mathbf{q}_{v_1},\dots,\mathbf{q}_{v_l} \rangle$.
We adopt a layer-wise time embedding strategy, where the embedding of time step $t$ is added to the user representation $\mathbf{p}_u$.
The output $\mathbf{\hat{x}}_0$, formulated as $\langle \mathbf{\hat{p}}_u, \mathbf{\hat{q}}_{v_1},\dots,\mathbf{\hat{q}}_{v_l} \rangle$, is projected back to the discrete item space using cosine similarity with the item embedding matrix $\mathbf{Q}$. The $i$-th generated item is represented as a one-hot vector $\mathbf{\hat{x}}^{\mathbf{v}_i}$.
Additionally, the distance metric $d(\cdot)$ is defined using Kullback–Leibler divergence to encourage $F_{\mathbf{\theta}}$ to produce consistent outputs across different partially masked historical interactions of users.

\renewcommand{\thealgorithm}{1}
\begin{algorithm}[htp]
\caption{Multistep Sampling Process} 
\begin{algorithmic}
    \Require 
    $F_{\mathbf{\theta}}$: Consistency function, $\mathbf{x}_{T}$: Absorbing state initialization, $T$: Total diffusion step, $[{\delta}_1, {\delta}_2 \dots {\delta}_N = T]$: Discretized diffusion steps, $N$: Sampling steps.
    \State $\mathbf{\hat{x}} \leftarrow F_{\mathbf{\theta}}(\mathbf{x}_T, T) $
    \For{$n = N-1$ to $0$} 
        \State $\mathbf{x}_{\delta_n} \sim q_{t|0}(\cdot| \mathbf{\hat{x}}^{\mathbf{v}_i}) = \text{exp} (\overline{\beta}(\delta_n)\mathbf{R})_{\mathbf{\hat{x}}^{\mathbf{v}_i} \cdot}$
        \State $\mathbf{\hat{x}} \leftarrow F_{\mathbf{\theta}}(\mathbf{x}_{\delta_n}, \delta_n) $
    
    \EndFor
    \Ensure
    $\mathbf{\hat{x}}$: Prediction of user-item interaction
\end{algorithmic}
\label{alg:sampling}
\end{algorithm}

\subsection{Contrastive Guided Diffusion}
Although discrete-space diffusion algorithms demonstrate strong capability in modeling data distributions, these methods primarily focus on learning direct historical interactions (i.e., one-hop neighbors in collaborative graphs) while neglecting valuable multi-hop relational information. To bridge this gap, CDRec incorporates contrastive learning to guide the diffusion process with structural information encoded in the user representations $\mathbf{P}$.

\subsubsection{Contrastive Learning.}
Given the generated interaction sequence, $\langle \mathbf{\hat{v}}_1, \mathbf{\hat{v}}_{2}, \dots, \mathbf{\hat{v}}_{l} \rangle$, we aggregate the corresponding collaborative embeddings to construct the user representation $\mathbf{e}_u$, simulating the propagation behavior of a single GNN layer ~\cite{zhang2023contrastive, lee2025scone, liu2025glprotein}. 
To effectively transfer domain knowledge from the collaborative graph, we utilize the pre-trained user embedding $\mathbf{p}_u$ as a positive sample while employing a randomly sampled non-interacted item set $\mathcal{V}^-_u$ as negative samples for contrastive learning. 
The CDRec employs the InfoNCE loss as our self-supervised learning objective to guide the diffusion generation, which can be defined as:
\begin{equation}
    \mathcal{L}^{cl} = \sum_{u \in \mathcal{U}} -\log \frac{\varsigma(\mathbf{e}_u, \mathbf{p}_{u})}{\varsigma(\mathbf{e}_u, \mathbf{p}_{u}) + \sum\limits_{v \in \mathcal{V}^-_u} \varsigma(\mathbf{e}_u, \mathbf{q}_{v})) },
    \label{eq:NCEloss}
\end{equation}
where the similarity function is defined as $\varsigma(\mathbf{\hat{e}}_{u}, \mathbf{p}_{*}) = \exp\left({\cos(\mathbf{\hat{e}}_{u}, \mathbf{p}_{*})} / \tau \right)$ to compute the cosine similarity between embeddings, and $\tau$ is the temperature parameter.

\subsubsection{Joint Training Mechanism.}
The joint optimization objective for CDRec combines multiple learning objectives through the following weighted formulation:
\begin{equation}
    \mathcal{L}= \lambda_1\mathcal{L}^{Con} + (1-\lambda_1) \mathcal{L}^{Diff} + \lambda_2 \mathcal{L}^{cl},
    \label{loss}
\end{equation}
where $\lambda_1$ balances two complementary objectives: the consistency loss, which emphasizes pattern consistency by producing identical predictions for interactions with varying masking levels, and the diffusion loss, which performs behavior reconstruction by directly recovering the original user–item interactions; while $\lambda_2$ balances the guidance from the structural information of the collaborative graph with the modeling of direct one-hop relations.
The complete training procedure for the denoising model is formally presented in Algorithm~\ref {alg:training}.

\renewcommand{\thealgorithm}{2}
\begin{algorithm}[thp]
\caption{Training Algorithm of CDRec} 
\begin{algorithmic}
    \Require $f_{\mathbf{\theta}}(\mathbf{x},t)$: Consistency function, $\nabla {t}$ represents the fixed time interval, $\text{\textbf{RecEnc}}(\cdot)$: Pre-trained GNN-based recommendation model.
    \Repeat
    \State Sample the historical interaction $\mathbf{x}_u = [\mathbf{v}_1, \mathbf{v}_2, \dots, \mathbf{v}_l]$ of user 
    \State $u$ as $\mathbf{x}_0$.
    \State $t_n \leftarrow t \sim \mathbf{\mathcal{U}}(\varepsilon, T)$ 
    \State $\mathbf{x}_{t_n} \sim q_{t_n|0}(\cdot| \mathbf{x}_0^{\mathbf{v}_i}) = \text{exp} (\overline{\beta}(t_n)\mathbf{R})_{\mathbf{x}_0^{\mathbf{v}_i} \cdot}$
    \State $t_{n-1} \leftarrow t_n - \nabla {t}$ 
    \If {Using One-step Recovery}
        \State Sample $\mathbf{\bar{x}}_{t_{n-1}}$ by recovering only one masked item with 
        \State the lowest absorbing probability.  
    \ElsIf {Using Pseudo-Euler}
        \State Sample $\mathbf{\bar{x}}_{t_{n-1}}$ via Equation ~\ref{eq:SDE_ins}.
    \EndIf
    \State Encode the $\mathbf{x}_{t_n}$ and $\mathbf{\bar{x}}_{t_{n-1}}$ with $\text{\textbf{RecEnc}}(\cdot)$.
    \State Compute the $\mathcal{L}^{Con}$ via Equation ~\ref{eq:consistency_loss}.
    \State Compute the $\mathcal{L}^{Diff}$ via Equation ~\ref{eq:diffusion_loss}.
    \State Sample non-interacted item set $\mathcal{V}^-_u$.
    \State Compute the $\mathcal{L}^{cl}$ via Equation ~\ref{eq:NCEloss}.
    \State Update the parameters $\mathbf{\theta}$ via gradient descent.
    \Until {converged}
    \Ensure
    $f_\mathbf{\theta}$: Optimized consistency function
\end{algorithmic}
\label{alg:training}
\end{algorithm}

%% file: sections/Experiment.tex
\section{Experiment}
\label{Experiment}

In this section, we conduct comprehensive experiments to assess the performance of our proposed CDRec framework.

\subsection{Experiment Settings}

\subsubsection{Datasets}
Experiments are conducted on three real-world datasets: Ciao ~\footnote{Ciao: \url{https://www.cse.msu.edu/~tangjili/trust.html}}, MovieLens-1M ~\footnote{MovieLens-1M: \url{https://grouplens.org/datasets/movielens/1m/}}, and Dianping ~\footnote{Dianping: \url{https://grouplens.org/datasets/movielens/1m/}}, where user-item ratings follow a $[1-5]$ scale. Following a widely adopted setting, we treat ratings of 3 and above as implicit interactions. Each dataset is partitioned into training, validation, and testing sets with an $8:1:1$ ratio. Dataset statistics are summarized in Table ~\ref {tab:dataset_statistics}.

\input{tables/data_statistics}

\subsubsection{Evaluation Metrics.}
We evaluate top-$K$ recommendation performance using two standard ranking metrics: Recall@$K$ (R@$K$) and Normalized Discounted Cumulative Gain (N@$K$), with $K\in\{5, 10\}$  under the full-ranking setting. All results are averaged over five independent runs to ensure reliability.

\input{tables/perf}

\subsubsection{Baselines.}
We evaluate the performance of CDRec against state-of-the-art baselines from three categories: GNN-based recommenders (NGCF~\cite{wang2019neural}, LightGCN~\cite {he2020lightgcn}, SGL~\cite{wu2021self}), VAE-based models (Mult-VAE~\cite{liang2018variational}, VGCL~\cite{yang2023generative}), and diffusion-based approaches (DiffRec~\cite{wang2023diffusion}, DiffRec-L~\cite{wang2023diffusion}, BSPM~\cite{choi2023blurring}, GiffCF~\cite{zhu2024graph}). Descriptions of these baselines are provided below:

\begin{itemize}[leftmargin=*]
\item NGCF ~\cite{wang2019neural}: NGCF employs graph convolutional networks to model collaborative signals through embedding propagation. 
\item LightGCN ~\cite{he2020lightgcn}: LightGCN simplifies NGCF by eliminating feature transformation and nonlinear activation layers, focusing specifically on collaborative filtering tasks.
\item SGL~\cite{wu2021self}: SGL introduces multiple auxiliary tasks to capture structural relatedness across different views, leveraging this information to enhance the supervision signals.
\item Mult-VAE~\cite{liang2018variational}: Mult-VAE produces interaction vectors via variational inference based on a multinomial distribution.
\item VGCL~\cite{yang2023generative}: VGCL integrates VAE with contrastive learning by employing variational graph reconstruction to generate multiple views for collaborative filtering.
\item DiffRec~\cite{wang2023diffusion}: DiffRec first introduces the diffusion algorithm into recommendation, perturbing the user’s historical interactions across the entire item vector.
\item L-DiffRec~\cite{wang2023diffusion}: L-DiffRec encodes collaborative data using a VAE framework and performs the diffusion process in the latent space.
\item BSPM~\cite{choi2023blurring}: BSPM leverages score-based generative models for collaborative filtering by incorporating a blurring and sharpening process on the interaction matrix.
\item GiffCF~\cite{zhu2024graph}: GiffCF applies a diffusion algorithm, grounded in the heat equation, to the item–item similarity graph.
\end{itemize}

\subsubsection{Parameter Settings}
The proposed CDRec framework is implemented in PyTorch. We set the batch size to 1024 for all three datasets, with the sequence length $l$ set to 20. CDRec is optimized using the Adam optimizer, and the learning rate is tuned from $\{0.001, 0.0001, 0.0001\}$. 
The number of sampling steps $N$ is selected from $\{1,3,5,10,30,50,100\}$, with $T$ is evaluated in the range $\{ N, 2N\}$ and the corresponding step size adjusted within $\nabla t$ within $\{1,5,10,20\}$. 
The hyperparameter $\lambda_1$ is searched in the range $[0.1,1]$ while the $\lambda_2$ is selected from $\{1.0, .01, 0.001, 0.0001\}$. 
We empirically set the $\omega = 0.5$, and $\sigma = T/10$.
In this work, user and item representations are initialized using a pre-trained 3-layer LightGCN model.

\subsection{Overall Performance}
We conduct a comprehensive evaluation of recommendation performance by comparing the proposed CDRec framework against existing baselines. As shown in Table~\ref {OverallPerfromance}, we report Recall and NDCG metrics across three benchmark datasets: Ciao, MovieLens-1M, and Dianping. Our main findings include:

\begin{itemize}[leftmargin=*]

\item Generative recommender systems outperform GNN-based approaches, particularly on large-scale sparse datasets, by directly modeling user interaction distributions rather than learning low-rank embeddings from high-dimensional interaction matrices.

\item While diffusion models outperform baseline approaches by capturing complex data distributions more effectively, their performance improvements remain modest, likely due to information loss during discrete-to-continuous data transformation.

\item CDRec demonstrates consistent superiority over all baseline methods across all evaluated datasets. The performance gains can be attributed to three key components.
Our popularity-aware noise scheduler effectively simulates real-world interaction patterns through an easy-to-hard sampling process.
The proposed consistency function accurately captures interaction distributions conditioned on the collaborative signals, leading to substantially improved recommendation accuracy.
Moreover, the contrastive learning component bridges the information gap between one-hop and multi-hop neighborhood representations in the collaborative graph, enhancing personalized sampling.
\end{itemize}

\begin{figure}[htp]
    \centering
    \begin{subfigure}[b]{0.157\textwidth}
        \includegraphics[width=\textwidth]{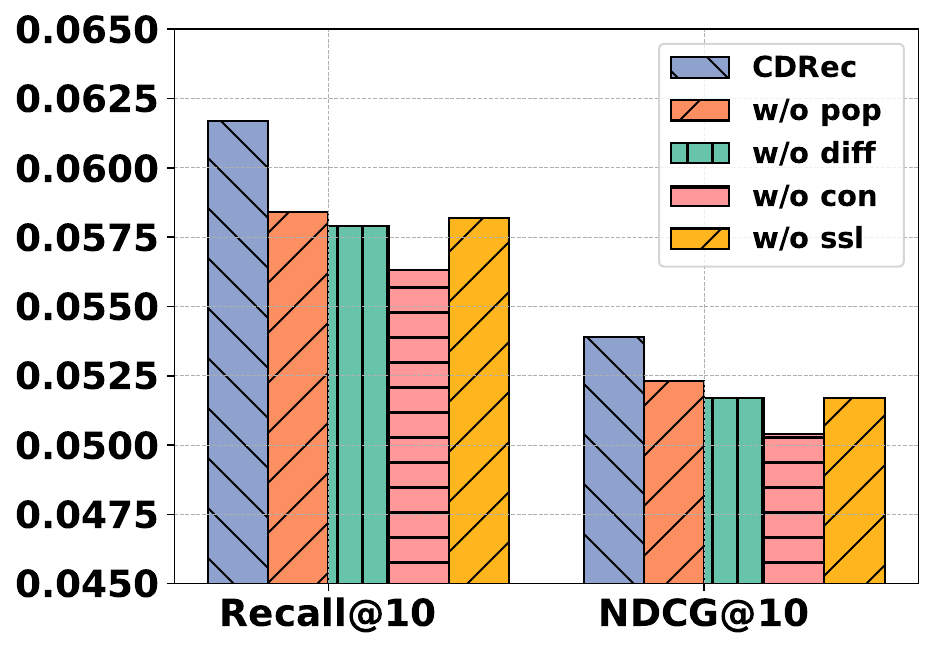}
        \caption{Ciao}
    \end{subfigure}
    \begin{subfigure}[b]{0.15\textwidth}
        \includegraphics[width=\textwidth]{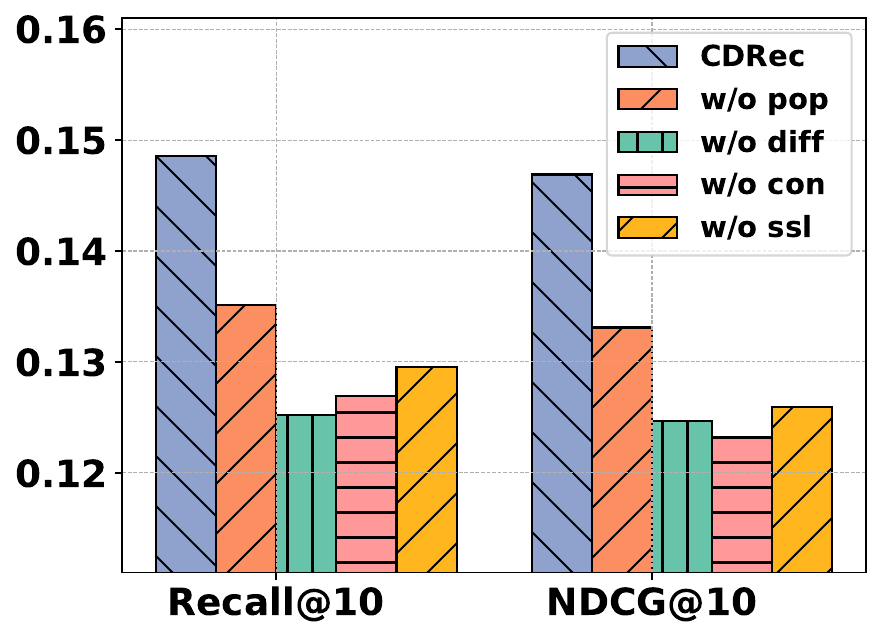}
        \caption{MovieLens-1M}
    \end{subfigure}
    \begin{subfigure}[b]{0.157\textwidth}
        \includegraphics[width=\textwidth]{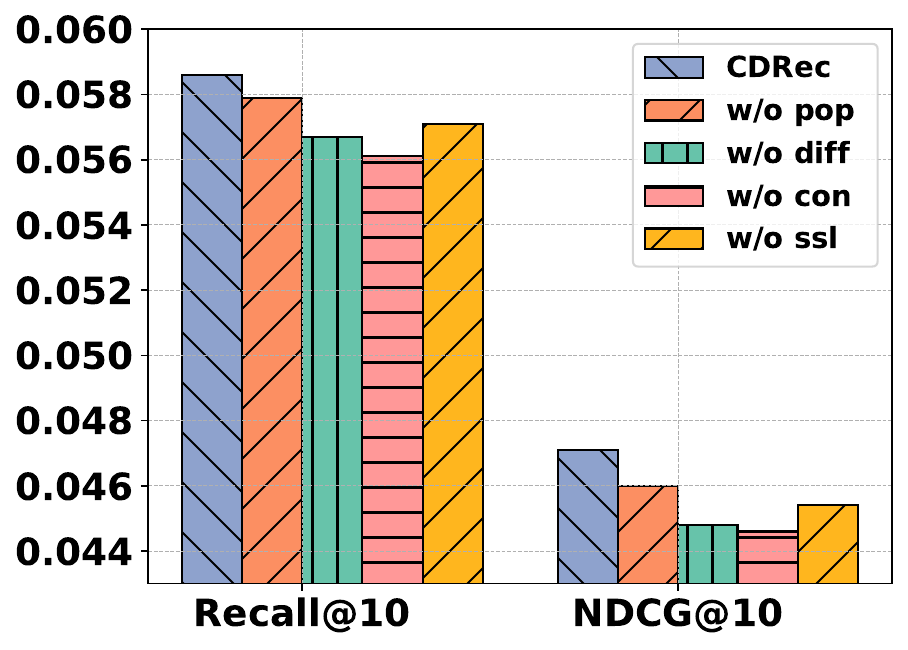}
        \caption{Dianping}
    \end{subfigure}
    \caption{Ablation study of CDRec and its variants on three datasets with evaluation metrics Recall@10 and NDCG@10.}
    \label{fig:ablation}
    \vskip -0.2in
\end{figure}

\subsection{Ablation Study}
To systematically evaluate the core components of CDRec, we design four architectural variants:

\begin{itemize}[leftmargin=*]
\item \textit{w/o pop}: Ablates the popularity-aware component, implementing standard uniform masking with linear noise schedule.
\item \textit{w/o con}: Removes the consistency loss $\mathcal{L}^{con}$, forcing the model to estimate reverse densities $p_{0|t}$ in the discrete state space.
\item \textit{w/o diff}: Eliminate the diffusion loss $\mathcal{L}^{diff}$, relying solely on the consistency loss for interaction distribution learning.
\item \textit{w/o ssl}: Ablates the contrastive learning objective $\mathcal{L}^{cl}$, resulting in a denoising process that is solely conditioned on the user representation prefix.
\end{itemize}

Figure~\ref{fig:ablation} presents our ablation study results. The popularity-aware noise scheduler proves significantly more effective than linear scheduling, demonstrating its ability to capture real-world interaction patterns while improving training stability and sampling quality. Performance suffers most severely when removing either the consistency or diffusion losses, confirming their fundamental role in modeling interaction distributions. Additionally, the contrastive learning component contributes measurably to performance, verifying its effectiveness in exploiting multi-hop relational patterns for personalized recommendation generation.

\begin{figure}[bp]
    \centering
    \vskip -0.15in
    \begin{subfigure}[b]{0.156\textwidth}
        \includegraphics[width=\textwidth]{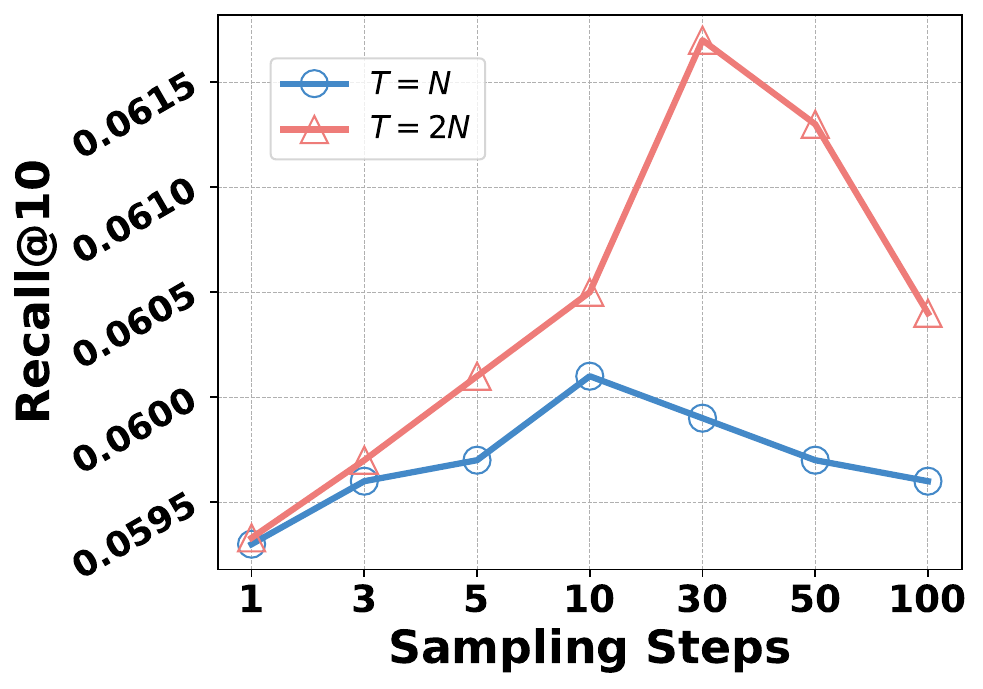}
        \caption{Ciao}
    \end{subfigure}
    \begin{subfigure}[b]{0.156\textwidth}
        \includegraphics[width=\textwidth]{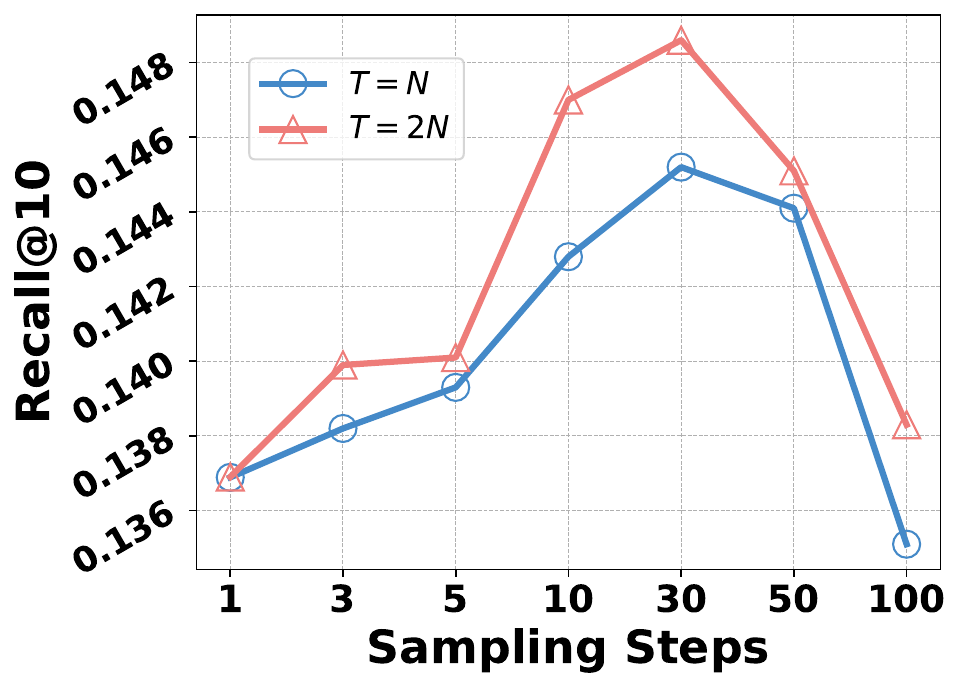}
        \caption{MovieLens-1M}
    \end{subfigure}
    \begin{subfigure}[b]{0.156\textwidth}
        \includegraphics[width=\textwidth]{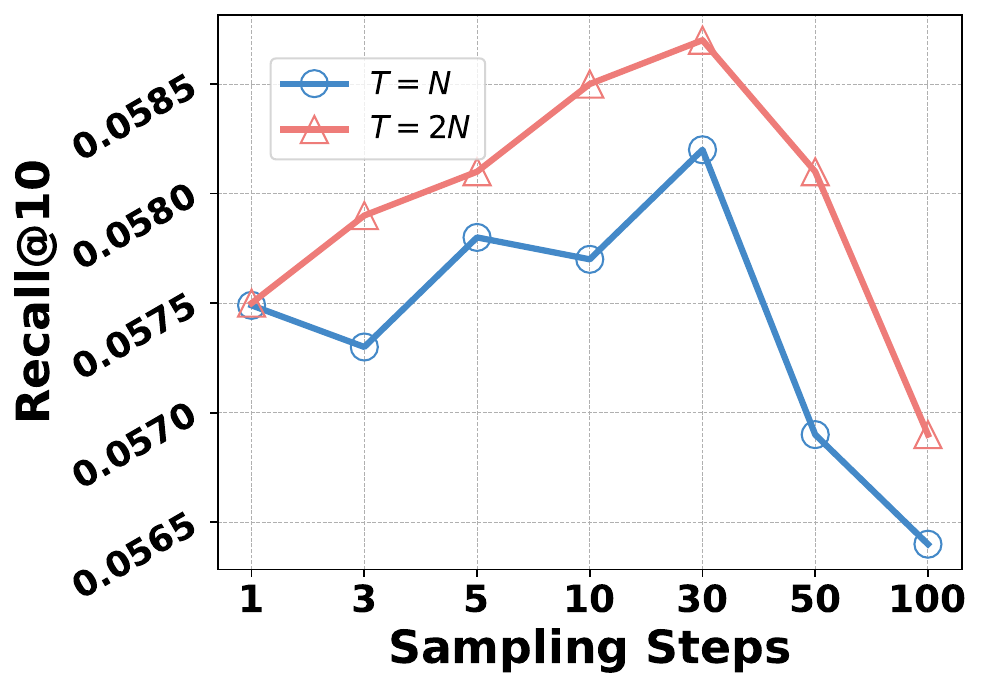}
        \caption{Dianping}
    \end{subfigure}
    
    \begin{subfigure}[b]{0.156\textwidth}
        \includegraphics[width=\textwidth]{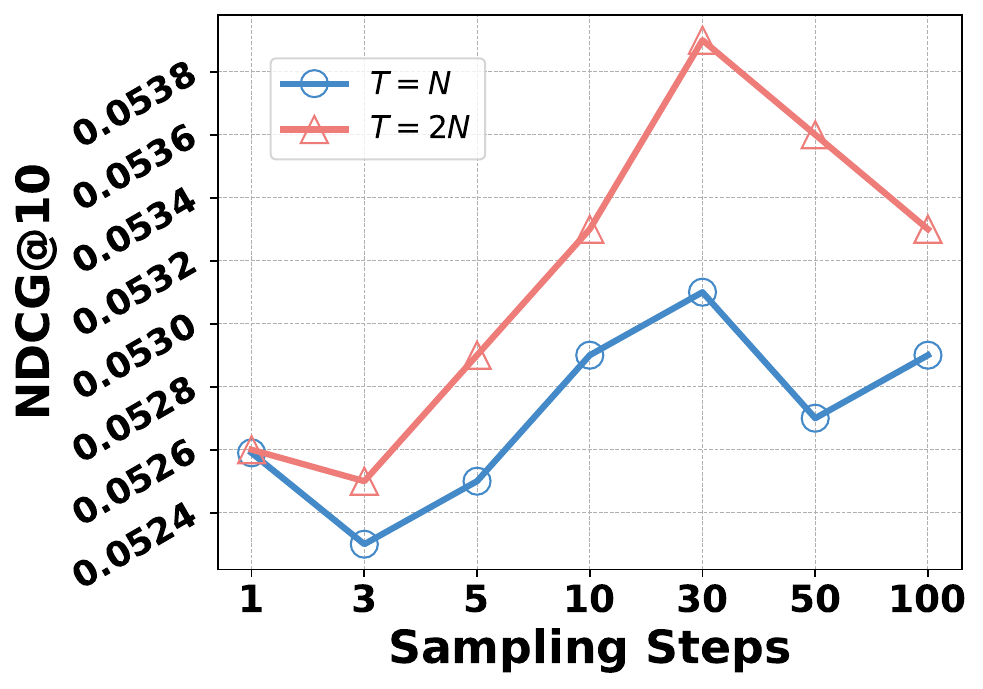}
        \caption{Ciao}
    \end{subfigure}
    \begin{subfigure}[b]{0.156\textwidth}
        \includegraphics[width=\textwidth]{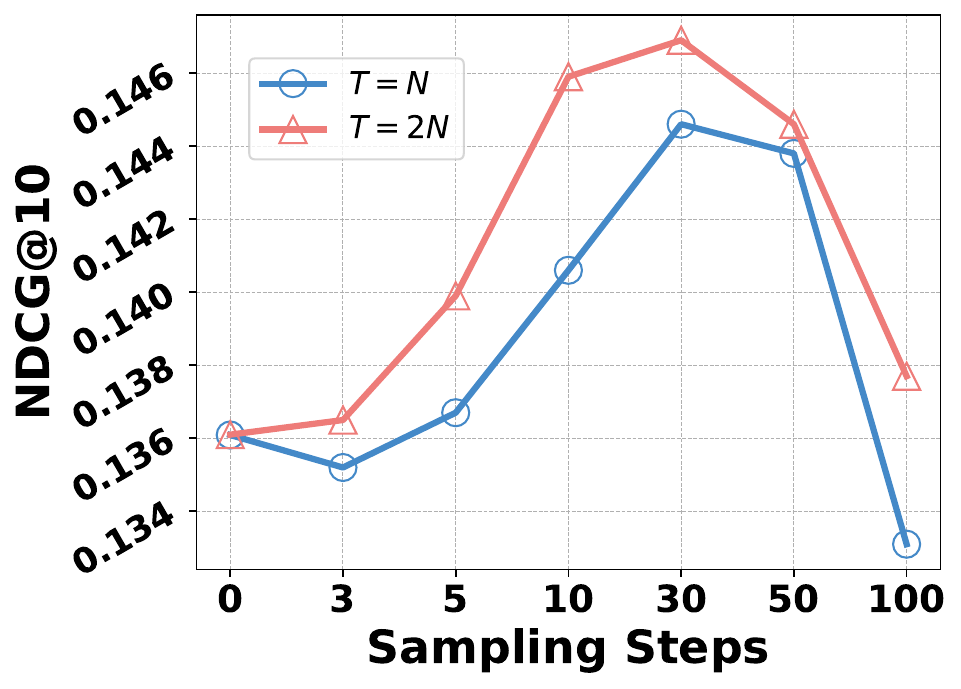}
        \caption{MovieLens-1M}
    \end{subfigure}
    \begin{subfigure}[b]{0.156\textwidth}
        \includegraphics[width=\textwidth]{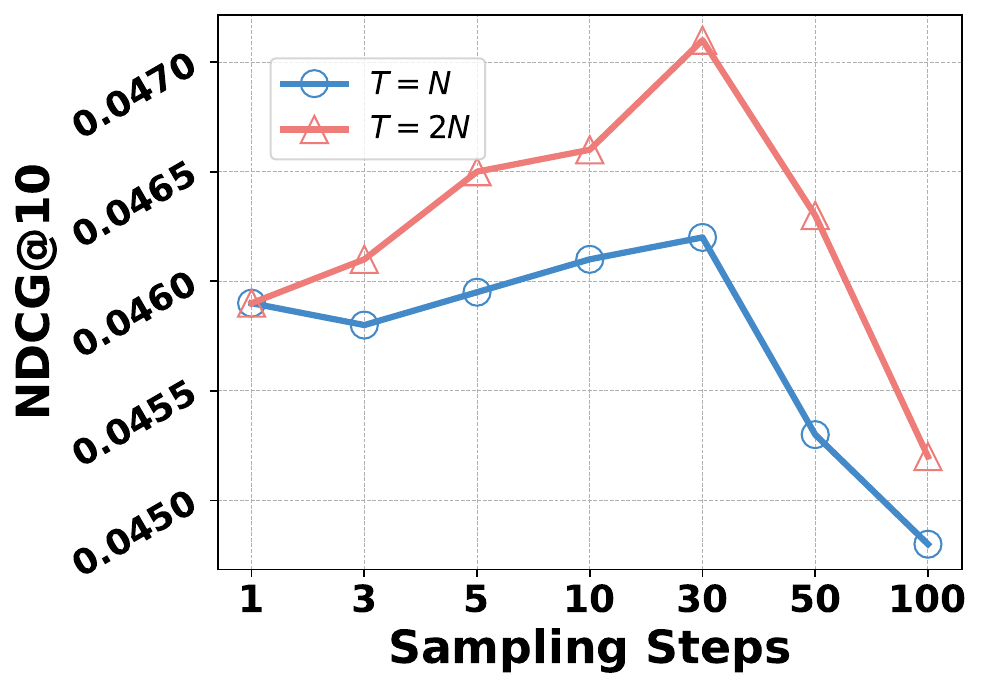}
        \caption{Dianping}
    \end{subfigure}
    \caption{Effect of sampling steps $N$ on Recall@10 and NDCG@10.
    }
    \label{fig:steps}
    \vskip -0.15in
\end{figure}

\subsection{Parameter Sensitivity Study}
In this section, we examine the impact of CDRec's key parameters.

\subsubsection{Diffusion Steps}

We evaluate the sampling steps $N$, with corresponding diffusion steps $T \in \{T= N, T =2N\}$. As shown in Figure~\ref{fig:steps}, optimal performance occurs at $30$ sampling steps with $T=60$ across all datasets. After this point, performance degrades with additional steps, likely due to noise accumulation from excessive sampling iterations. 
Moreover, experimental results indicate that larger $T$ values generally improve performance with the same sampling step, suggesting that an extended step $T$ better simulates the continuous-time diffusion trajectory and encourages the model to learn the interaction distribution.
Importantly, \textit{the one-step recommendation of CDRec achieves performance comparable to diffusion-based baselines, demonstrating the efficiency of the consistency parameterization}.

\subsubsection{Loss Weights}

\begin{figure}[tp]
    \centering
    \begin{subfigure}[b]{0.155\textwidth}
        \includegraphics[width=\textwidth]{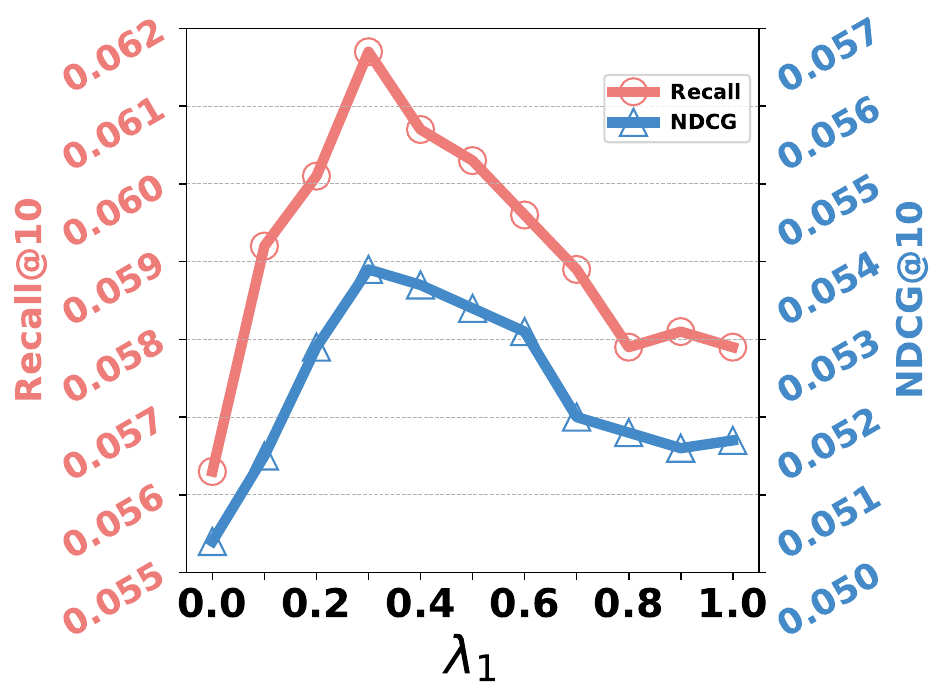}
        \caption{Ciao}
    \end{subfigure}
    \begin{subfigure}[b]{0.155\textwidth}
        \includegraphics[width=\textwidth]{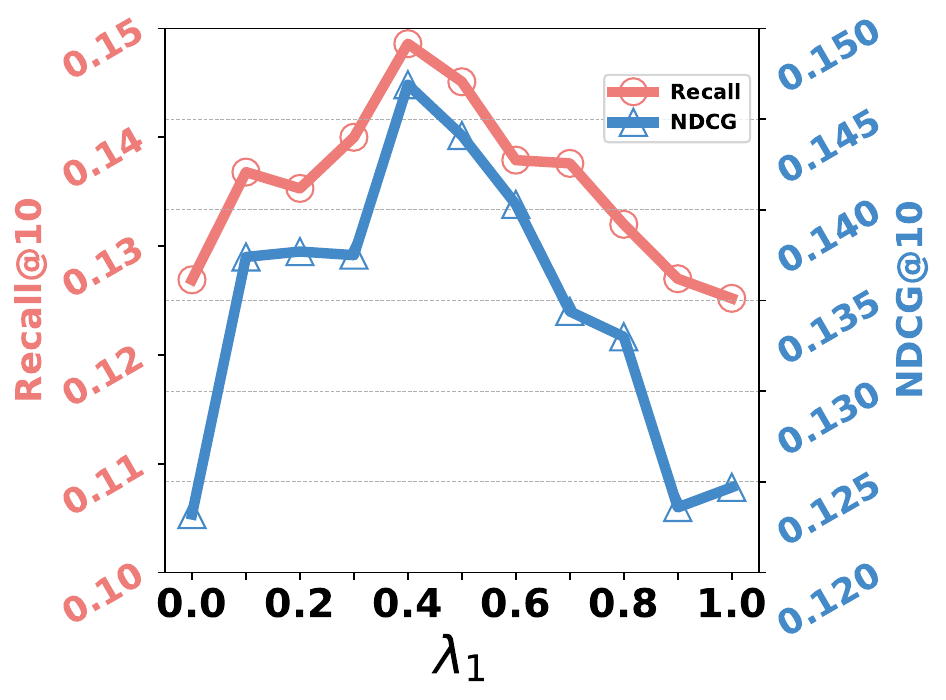}
        \caption{MovieLens-1M}
    \end{subfigure}
    \begin{subfigure}[b]{0.155\textwidth}
        \includegraphics[width=\textwidth]{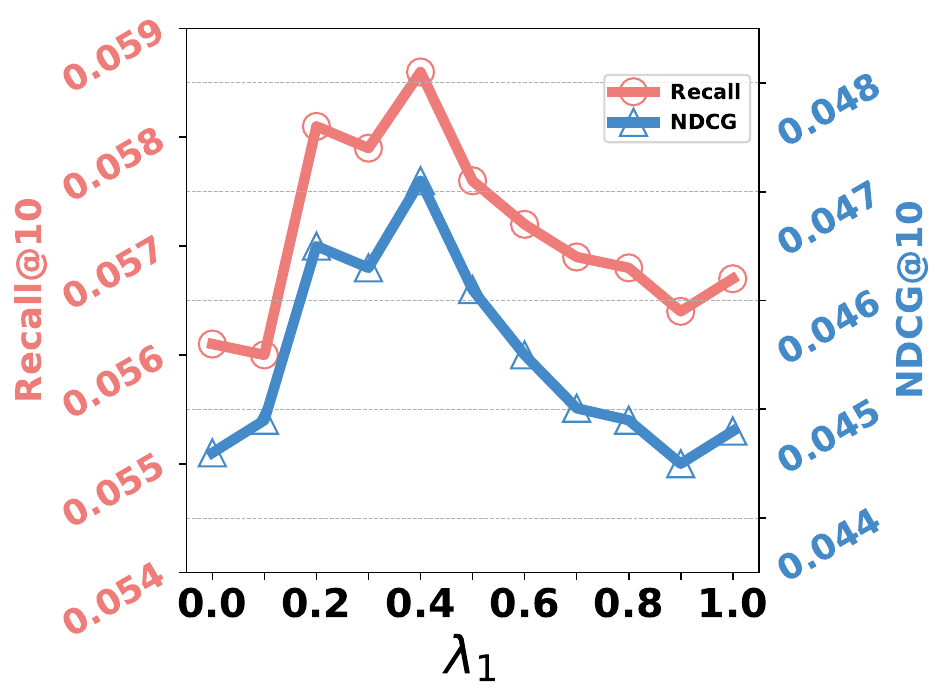}
        \caption{Dianping}
    \end{subfigure}
    \caption{Effect of $\lambda_1$ on Recall@10 and NDCG@10.
    }
    \label{fig:lambda}
    \vskip -0.13in
\end{figure}
We conduct the sensitivity study on $\lambda_1$ to examine CDRec's training dynamics, with results shown in Figure~\ref{fig:lambda}. The best performance is achieved at $\lambda_1 = 0.4$, with a decline observed as $\lambda_1$ deviates in either direction. These results suggest that incorporating $\mathcal{L}^{Diff}$ effectively enhances the modeling of interaction distributions in the recommendation context while mitigating undesired bias.

\subsubsection{Construction of $(\mathbf{\bar{x}}_{t_{n-1}}, \mathbf{x}_{t_n})$.}
\begin{figure}[thp]
    \centering
    \begin{subfigure}[b]{0.157\textwidth}
        \includegraphics[width=\textwidth]{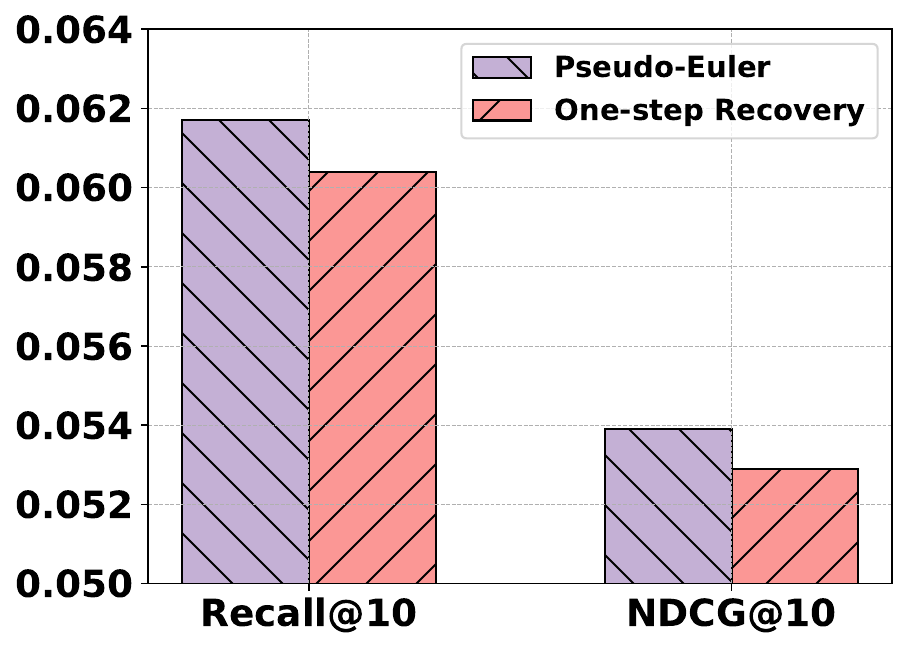}
        \caption{Ciao}
    \end{subfigure}
    \begin{subfigure}[b]{0.15\textwidth}
        \includegraphics[width=\textwidth]{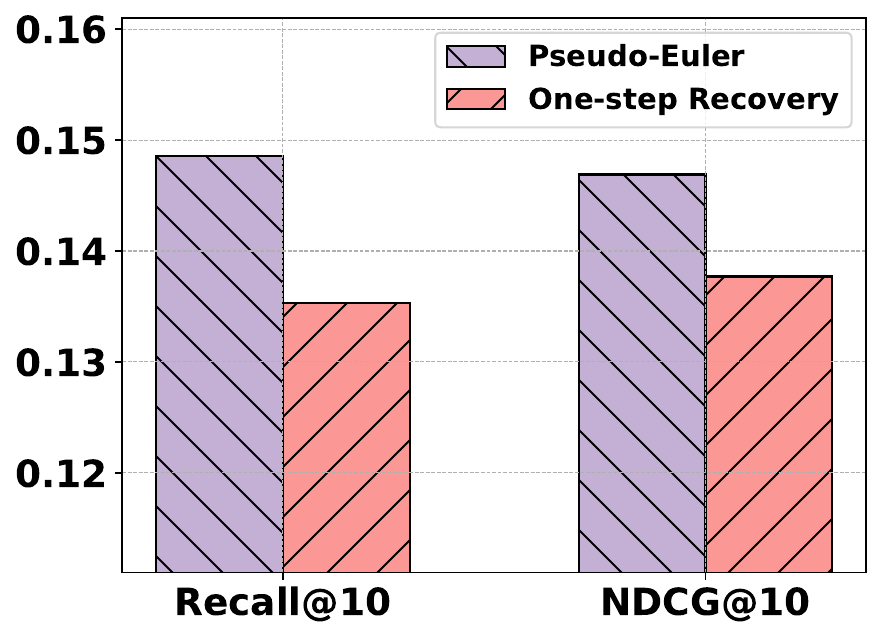}
        \caption{MovieLens-1M}
    \end{subfigure}
    \begin{subfigure}[b]{0.157\textwidth}
        \includegraphics[width=\textwidth]{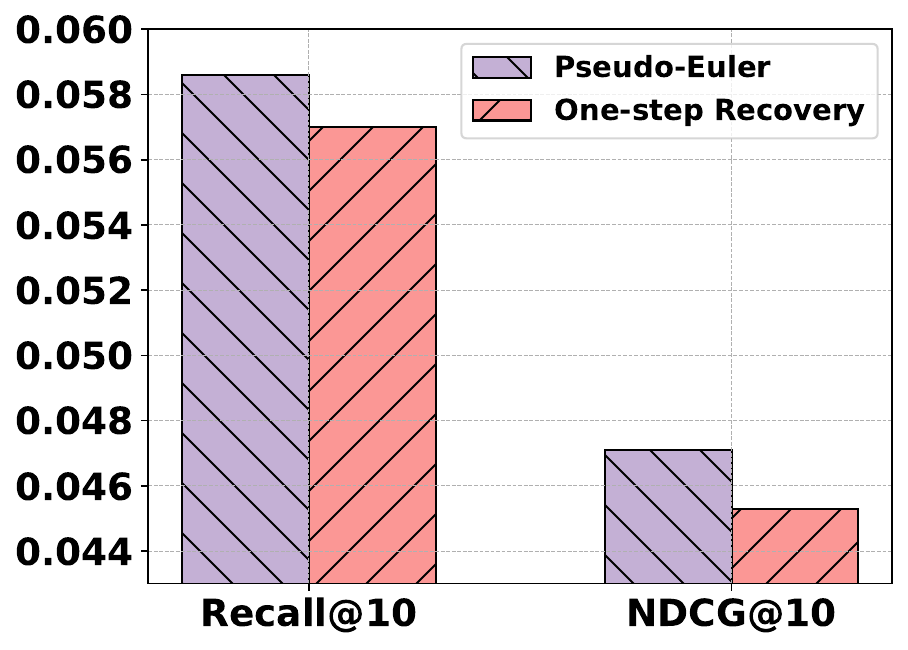}
        \caption{Dianping}
    \end{subfigure}
    \caption{Effect of methods to construct $(\mathbf{\bar{x}}_{t_{n-1}}, \mathbf{x}_{t_n})$ on  Recall@10 and NDCG@10.}
    \label{fig:construct}
    \vskip -0.2in
\end{figure}

We evaluate two methods, pseudo-Euler and one-step recovery, for constructing the data pair $(\mathbf{\bar{x}}_{t_{n-1}}, \mathbf{x}_{t_n})$ to train the consistency model $F_{\mathbf{\theta}}$. The results, shown in Figure ~\ref{fig:construct}, indicate that pseudo-Euler consistently outperforms one-step recovery across all three datasets. This advantage likely arises because Equation~\ref{eq:SDE_ins} adaptively demasks items based on time intervals $\nabla t$, thereby more accurately simulating the diffusion process.

\subsection{Model Investigation}
This section presents a comprehensive evaluation of CDRec, including its time efficiency and case study analyses.

\begin{figure}[tp]
    
    \centering
    \begin{subfigure}[t]{0.23\textwidth}
        \includegraphics[width=\textwidth]{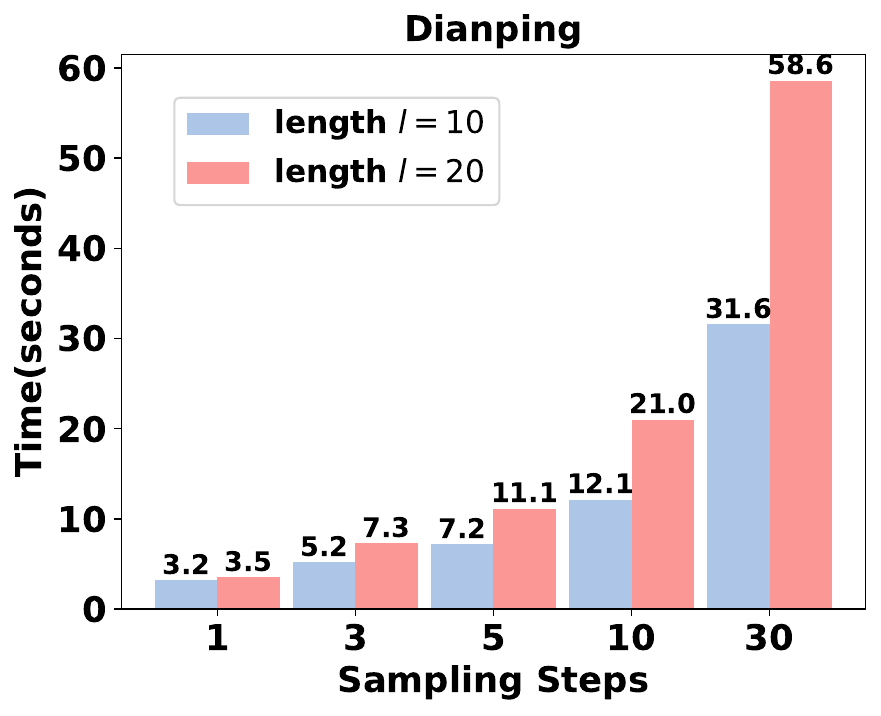}
        \caption{Sequence Length}
    \end{subfigure}
    \hfill
    \begin{subfigure}[t]{0.235\textwidth}
        \includegraphics[width=\textwidth]{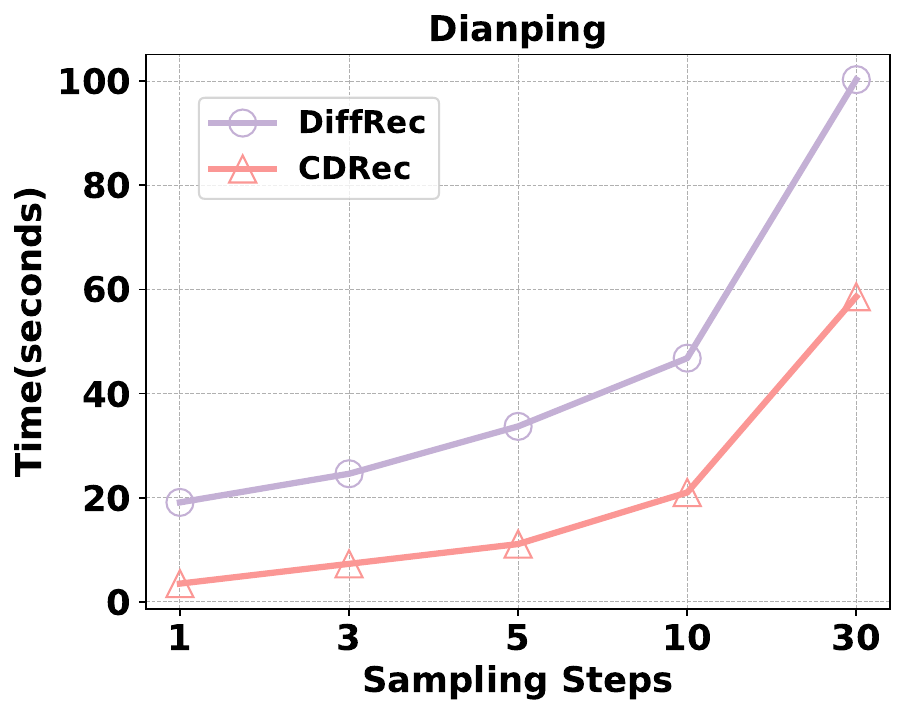}
        \caption{CDRec vs. DiffRec}
    \end{subfigure}
    \caption{Time Analysis}
    \label{fig:time}
    \vskip -0.12in
\end{figure}

\subsubsection{Time Analysis.}
We systematically evaluate the sampling efficiency of CDRec under varying parameter settings and conduct comparative analysis with the baseline DiffRec model. 
Figure~\ref{fig:time}(a) illustrates the time cost with respect to two key parameters: sequence length and sampling steps. The results show that time cost increases sharply with the number of sampling steps, underscoring the importance of achieving the single-step recommendation while maintaining comparable performance.
Figure~\ref{fig:time}(b) demonstrates the comparison of sampling time of CDRec and DiffRec on the Dianping dataset under the same settings (test batch size is 1024). CDRec outperforms DiffRec as its discrete diffusion algorithm perturbs data through masking operations rather than applying Gaussian noise to the entire item vector.

\subsubsection{Case Study.}
We illustrate the forward masking diffusion process for a specific user in the MovieLens-1M dataset across the diffusion steps. The observations are as follows: (1) The user’s preferences span action films, comedies, and dramas, with relatively popular films in these categories typically watched first, aligning with our assumption of user behavior. (2) The proposed popularity deviation metric guides the forward diffusion process, allowing more popular movies to be retained longer. However, the diffusion mechanism introduces randomness, so the masking order does not strictly follow interaction frequency. This randomness gives the model a greater opportunity to model less popular items, helping to mitigate potential bias.

\begin{figure}[thp]
\centering
\includegraphics[width=0.48\textwidth]{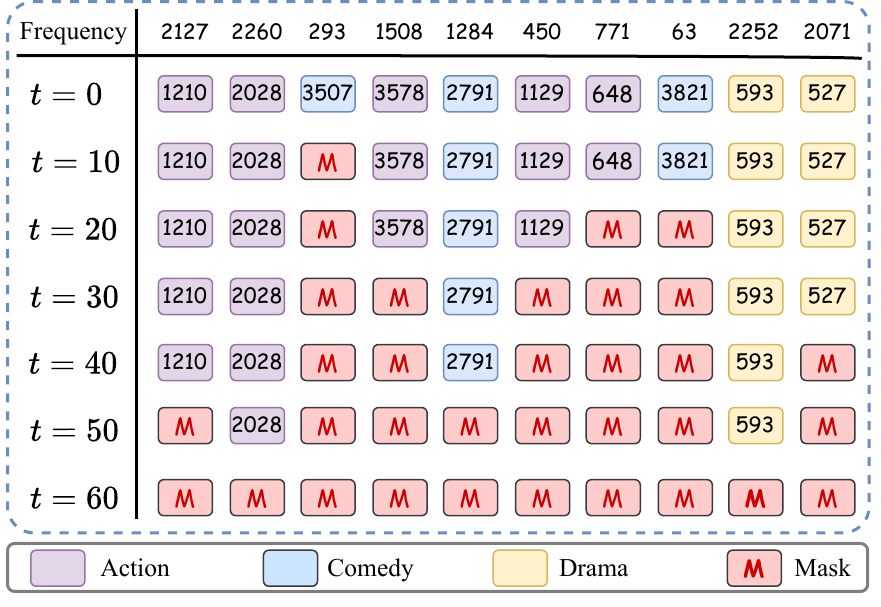}%
\caption{Case study of the forward diffusion process for user ID 98 in MovieLens-1M with $\omega=0.5$. Movies are initially ordered by interaction timestamps. }
\label{fig:case_study} 
\vskip -0.2in
\end{figure}

%% file: tables/data_statistics.tex
\begin{table}[htp]
\centering
\caption{Statistics of the datasets}
\begin{tabular}{c|ccc}
\toprule
\toprule 
\textbf{Dataset} & \textbf{Ciao} & \textbf{MovieLens-1M} & \textbf{Dianping}\\
\hline
\#Users & 7,373 & 6,038 & 131,629\\
\#Items & 91,091 & 3,533 & 10,932 \\
\#Interactions & 227,392 & 575,281 & 1,412,544 \\
Interaction Density & 0.0338\%  & 2.6967\% & 0.0981\% \\
\bottomrule
\bottomrule

\end{tabular}
\label{tab:dataset_statistics}
\vskip -0.15in
\end{table}

%% file: tables/perf.tex
\begin{table*}[htp]
\caption{Comparison of overall performance on three recommendation datasets. The best performance is highlighted in bold, and the second best is underlined. \%Improve denotes the relative improvement of CDRec over the strongest baseline.}
\vskip -0.1in
\begin{tabular}{c|cccc|cccc|cccc}
\toprule
\toprule
\textbf{Dataset}  & \multicolumn{4}{|c|}{\textbf{Ciao}} & \multicolumn{4}{c}{\textbf{MovieLens-1M}} & \multicolumn{4}{|c}{\textbf{Dianping}} \\
\cmidrule{1-13}
\textbf{Method}   & \textbf{R@10} & \textbf{R@5} & \textbf{N@10} & \textbf{N@5} & \textbf{R@10} & \textbf{R@5} & \textbf{N@10} & \textbf{N@5} & \textbf{R@10} & \textbf{R@5} & \textbf{N@10} & \textbf{N@5} \\
\hline
NGCF &  0.0483 & 0.0266 & 0.0396 & 0.0321 & 0.1201 & 0.0724 & 0.1212 & 0.1136 & 0.0531 & 0.0316 & 0.0381 & 0.0317\\
LightGCN & 0.0525 & 0.0312 & 0.0441 & 0.0370 & 0.1297 & 0.0783 & 0.1231 & 0.1193 & 0.0543 & 0.0333 & 0.0421 & 0.0339\\
SGL  & 0.0553 & 0.0364 & 0.0493 & 0.0438 & 0.1319 & 0.0852 & 0.1277 & 0.1207 & 0.0546 & 0.0339 & 0.0433 & 0.0365\\
Muti-VAE & 0.0517 & 0.0361 & 0.0409 & 0.0341 & 0.1337 & 0.0863 & 0.1193 & 0.1152 & 0.0547 & 0.0341 & 0.0415 & 0.0345\\
VGCL & 0.0568 & 0.0346 & 0.0477 & 0.0422 & 0.1339 & 0.0846 & 0.1315 & 0.1273 & 0.0551 & 0.0336 & 0.0431 & 0.0353\\
DiffRec & 0.0436 & 0.0289 & 0.0397 & 0.0361 & 0.1351 & 0.0839 & 0.1379 &\underline{0.1309} & 0.0556 & \underline{0.0345} & 0.0439 & 0.0367\\
L-DiffRec & 0.0458 & 0.0293 & 0.0417 & 0.0371 & 0.1326 & 0.0799 & 0.1351 & 0.1169 & 0.0539 & 0.0327 & 0.0435 & \underline{0.0371}\\
BSPM  & \underline{0.0590} & 0.0363 & \underline{0.0518} & \underline{0.0447} & 0.1399 & \underline{0.0871} &0.01347 & 0.1301 & \underline{0.0557} & 0.0329 & \underline{0.0441} & 0.0359\\
GiffCF & 0.0584 & \underline{0.0372} & 0.0473 & 0.0396 & \underline{0.1426} & 0.0857 & \underline{0.1395} & 0.1293 & 0.0505 &0.0339 & 0.0412 & 0.0361\\
\hline
CDRec &  \textbf{0.0617} & \textbf{0.0390} & \textbf{0.0539} & \textbf{0.0473} & \textbf{0.1486} & \textbf{0.0931} & \textbf{0.1469} & \textbf{0.1384} & \textbf{0.0586} & \textbf{0.0355}& \textbf{0.0471} & \textbf{0.0394} \\
\%Improve & 4.58\% & 4.84\% & 4.06\% & 5.81\% & 4.21\% & 6.88\% & 5.30\% & 5.72\% & 5.21\% & 2.90\% & 6.80\% & 6.19\% \\
\bottomrule
\bottomrule
\end{tabular}

\label{OverallPerfromance}
\end{table*}

%% file: sections/RelatedWork.tex
\section{Related Work}
\label{RelatedWork}

This section reviews the related work in diffusion models for recommender systems and discrete-space diffusion algorithms.

\subsection{Diffusion Recommender Model}
Diffusion models have emerged as a powerful generative approach for RS, achieving remarkable performance by modeling complex data  ~\cite{fan2023generative}.
In collaborative filtering, DiffRec~\cite{wang2023diffusion} pioneered diffusion-based recommendation by representing historical interactions as binary vectors and employing DDPM~\cite{ho2020denoising} to generate missing user-item interactions in the latent space.
Building on this, GiffCF~\cite{zhu2024graph} enhances the approach by integrating graph-level diffusion to capture higher-order relational patterns, while BSPM~\cite{choi2023blurring} adopts a continuous-time diffusion framework over the adjacency matrix.
On the other hand, DreamRec~\cite{yang2023generate} adopts the learn-to-generate paradigm, applying diffusion processes to target representations conditioned on historical interactions.
PreferDiff ~\cite{liu2024preference} introduces negative samples during training for sequential recommendation, potentially aligning with the Direct Preference Optimization (DPO) framework.
Despite their advancements, these methods require encoding discrete collaborative graph data into continuous space, which poses a risk of collaborative information loss.

DCDR~\cite{lin2024discrete} initially introduced the discrete-state diffusion algorithm into recommendation for the reranking task, serving as an early attempt. However, the discrete diffusion is only used to perturb item positions within interactions, without fully exploiting the modeling capacity of discrete diffusion frameworks. 
This highlights the necessity of developing a discrete diffusion algorithm that directly models the distribution of user preferences while enabling an efficient sampling process.

\subsection{Discrete Diffusion Model}
Discrete diffusion models have attracted growing attention due to their capacity to model complex distributions in discrete data spaces ~\cite{xu2024discrete}.
D3PM~\cite{austin2021structured} proposed the first discrete-state diffusion framework for categorical random variables in discrete time. 
Instead of using Gaussian noise, D3PM introduces several stationary distributions, such as the uniform distribution (where the transition probability to any other state is uniform) and the absorbing state (where perturbation is applied by masking the token).
\citet{campbell2022continuous} initially describes the discrete-state diffusion algorithm with SDE formulations in continuous time to enhance the flexibility of the sampling strategy. However, its discrete parameterization of the reverse diffusion process limits empirical performance.
\citet{sun2022score} introduced the ratio matching algorithm to model marginal probabilities via maximum likelihood training, followed by methods such as CSM~\cite{meng2022concrete} and SEDD~\cite{lou2024discrete}, which proposed alternative score matching approaches. However, the parameterization methods remain underdeveloped, hindered by requirements for specialized architectures, constraints on positive probabilities, and high computational cost. 

%% file: sections/Conclusion.tex
\section{Conclusion}
\label{Conclusion}
In this paper, we propose CDRec, a framework that employs a discrete-state diffusion algorithm in continuous time to model the distribution of user–item interactions. CDRec leverages an absorbing state to perturb historical interactions via masking operations and learns a parameterized reverse process to generate personalized recommendations.
To better reflect real-world interaction generation, we design a popularity-aware noise schedule in which items with higher interaction frequencies are assigned lower absorption probabilities during the diffusion process, thereby enabling an easy-to-hard generation strategy for accurate preference modeling. For the reverse process, we parameterize it with a consistency function that captures user behavior patterns over masked historical interactions, addressing the limitations of existing parameterization approaches in the context of recommendation systems.
This framework balances sampling quality and efficiency by supporting both single-step and multi-step generation. Extensive experiments on three real-world datasets show that CDRec consistently outperforms existing methods, validating its effectiveness.

\section{Ethical Considerations}

In this section, we discuss potential ethical issues associated with the proposed CDRec framework. Although designed to enhance recommendation accuracy and efficiency, CDRec may also pose risks related to data privacy, fairness, and bias.

\textbf{Data Privacy.} Recommendation systems inherently process user behavioral data, which may contain sensitive personal information. Although our framework does not require explicit personal identifiers, improper data handling or insufficient anonymization could still expose users to privacy risks, including re-identification and unauthorized data misuse. To address these concerns, we adhere to strict data usage policies, such as enforcing access controls and encryption, to ensure compliance with privacy regulations (e.g., GDPR, CCPA). Furthermore, robust anonymization techniques, such as k-anonymity and tokenization, are applied to safeguard sensitive attributes. These measures collectively mitigate privacy risks and protect user information throughout the training process.

\textbf{Fairness and Bias.} Since the model is trained on historical user–item interaction data, it may inherently encode societal biases or skewed popularity distributions. Without appropriate intervention, these biases could be perpetuated or even amplified, leading to unequal item exposure or unfair treatment of certain user groups. To mitigate such risks, regular bias audits should be performed to monitor disparate impacts using established fairness metrics. These audits, combined with fairness-aware model adjustments, can help ensure equitable outcomes while preserving recommendation performance.

%% file: sample-base.bib
@String{Computing = "Computing" }

@String{Springer = "Springer-Verlag" }

@ArtifactSoftware{R,
    title = {R: A Language and Environment for Statistical Computing},
    author = {{R Core Team}},
    organization = {R Foundation for Statistical Computing},
    address = {Vienna, Austria},
    year = {2019},
    url = {https://www.R-project.org/},
}

@inproceedings{fan2019graph,
  title={Graph neural networks for social recommendation},
  author={Fan, Wenqi and Ma, Yao and Li, Qing and He, Yuan and Zhao, Eric and Tang, Jiliang and Yin, Dawei},
  booktitle={Proc. World Wide Web Conf.},
  pages={417--426},
  year={2019}
}

@article{zhao2024recommender,
  title={Recommender systems in the era of large language models (llms)},
  author={Zhao, Zihuai and Fan, Wenqi and Li, Jiatong and Liu, Yunqing and Mei, Xiaowei and Wang, Yiqi and Wen, Zhen and Wang, Fei and Zhao, Xiangyu and Tang, Jiliang and others},
  journal={IEEE Trans. Knowl. Data Eng.},
  year={2024},
  publisher={IEEE}
}

@inproceedings{he2020lightgcn,
  title={Lightgcn: Simplifying and powering graph convolution network for recommendation},
  author={He, Xiangnan and Deng, Kuan and Wang, Xiang and Li, Yan and Zhang, Yongdong and Wang, Meng},
  booktitle={Proc. 43rd Int. ACM SIGIR Conf. Res. Develop. Inf. Retrieval},
  pages={639--648},
  year={2020}
}

@inproceedings{wu2021self,
  title={Self-supervised graph learning for recommendation},
  author={Wu, Jiancan and Wang, Xiang and Feng, Fuli and He, Xiangnan and Chen, Liang and Lian, Jianxun and Xie, Xing},
  booktitle={Proc. 44th Int. ACM SIGIR Conf. Res. Develop. Inf. Retrieval},
  pages={726--735},
  year={2021}
}

@inproceedings{fan2022graph,
  title={Graph trend filtering networks for recommendation},
  author={Fan, Wenqi and Liu, Xiaorui and Jin, Wei and Zhao, Xiangyu and Tang, Jiliang and Li, Qing},
  booktitle={Proc. 45th Int. ACM SIGIR Conf. Res. Develop. Inf. Retrieval},
  pages={112--121},
  year={2022}
}

@article{yang2023generate,
  title={Generate what you prefer: Reshaping sequential recommendation via guided diffusion},
  author={Yang, Zhengyi and Wu, Jiancan and Wang, Zhicai and Wang, Xiang and Yuan, Yancheng and He, Xiangnan},
  journal={Proc. Adv. Neural Inf. Process. Syst.},
  volume={36},
  pages={24247--24261},
  year={2023}
}

@article{wang2023generative,
  title={Generative recommendation: Towards next-generation recommender paradigm},
  author={Wang, Wenjie and Lin, Xinyu and Feng, Fuli and He, Xiangnan and Chua, Tat-Seng},
  journal={arXiv preprint arXiv:2304.03516},
  year={2023}
}

@article{qu2025generative,
  title={Generative Recommendation with Continuous-Token Diffusion},
  author={Qu, Haohao and Fan, Wenqi and Lin, Shanru},
  journal={arXiv preprint arXiv:2504.12007},
  year={2025}
}

@inproceedings{wang2023diffusion,
  title={Diffusion recommender model},
  author={Wang, Wenjie and Xu, Yiyan and Feng, Fuli and Lin, Xinyu and He, Xiangnan and Chua, Tat-Seng},
  booktitle={Proc. 46th Int. ACM SIGIR Conf. Res. Develop. Inf. Retrieval},
  pages={832--841},
  year={2023}
}

@inproceedings{lin2024discrete,
  title={Discrete conditional diffusion for reranking in recommendation},
  author={Lin, Xiao and Chen, Xiaokai and Wang, Chenyang and Shu, Hantao and Song, Linfeng and Li, Biao and Jiang, Peng},
  booktitle={Companion Proceedings of the ACM Web Conference 2024},
  pages={161--169},
  year={2024}
}

@article{ho2020denoising,
  title={Denoising diffusion probabilistic models},
  author={Ho, Jonathan and Jain, Ajay and Abbeel, Pieter},
  journal={Proc. Adv. Neural Inf. Process. Syst.},
  volume={33},
  pages={6840--6851},
  year={2020}
}

@inproceedings{
  song2021scorebased,
  title={Score-Based Generative Modeling through Stochastic Differential Equations},
  author={Yang Song and Jascha Sohl-Dickstein and Diederik P Kingma and Abhishek Kumar and Stefano Ermon and Ben Poole},
  pages = {240-276},
  booktitle={Proc. 9th Int. Conf. Learn. Representations},
  year={2021},
}

@inproceedings{choi2023blurring,
  title={Blurring-sharpening process models for collaborative filtering},
  author={Choi, Jeongwhan and Hong, Seoyoung and Park, Noseong and Cho, Sung-Bae},
  booktitle={Proc. 46th Int. ACM SIGIR Conf. Res. Develop. Inf. Retrieval},
  pages={1096--1106},
  year={2023}
}

@inproceedings{zhao2024denoising,
  title={Denoising diffusion recommender model},
  author={Zhao, Jujia and Wenjie, Wang and Xu, Yiyan and Sun, Teng and Feng, Fuli and Chua, Tat-Seng},
  booktitle={Proc. 47th Int. ACM SIGIR Conf. Res. Develop. Inf. Retrieval},
  pages={1370--1379},
  year={2024}
}

@inproceedings{zhu2024graph,
  title={Graph signal diffusion model for collaborative filtering},
  author={Zhu, Yunqin and Wang, Chao and Zhang, Qi and Xiong, Hui},
  booktitle={Proc. 47th Int. ACM SIGIR Conf. Res. Develop. Inf. Retrieval},
  pages={1380--1390},
  year={2024}
}

@inproceedings{
    liu2024preference,
    title={Preference Diffusion for Recommendation},
    author={Shuo Liu and An Zhang and Guoqing Hu and Hong Qian and Tat-Seng Chua},
    booktitle={Proc. 13th Int. Conf. Learn. Representations},
    pages={11043--11081},
    year={2025},
}

@inproceedings{wang2019neural,
  title={Neural graph collaborative filtering},
  author={Wang, Xiang and He, Xiangnan and Wang, Meng and Feng, Fuli and Chua, Tat-Seng},
  booktitle={Proc. 42nd Int. ACM SIGIR Conf. Res. Develop. Inf. Retrieval},
  pages={165--174},
  year={2019}
}

@inproceedings{liang2018variational,
  title={Variational autoencoders for collaborative filtering},
  author={Liang, Dawen and Krishnan, Rahul G and Hoffman, Matthew D and Jebara, Tony},
  booktitle={Proc. World Wide Web Conf.},
  pages={689--698},
  year={2018}
}

@article{austin2021structured,
  title={Structured denoising diffusion models in discrete state-spaces},
  author={Austin, Jacob and Johnson, Daniel D and Ho, Jonathan and Tarlow, Daniel and Van Den Berg, Rianne},
  journal={Proc. Adv. Neural Inf. Process. Syst.},
  volume={34},
  pages={17981--17993},
  year={2021}
}

@article{xie2024breaking,
  title={Breaking determinism: Fuzzy modeling of sequential recommendation using discrete state space diffusion model},
  author={Xie, Wenjia and Wang, Hao and Zhang, Luankang and Zhou, Rui and Lian, Defu and Chen, Enhong},
  journal={Proc. Adv. Neural Inf. Process. Syst.},
  volume={37},
  pages={22720--22744},
  year={2024}
}

@article{campbell2022continuous,
  title={A continuous time framework for discrete denoising models},
  author={Campbell, Andrew and Benton, Joe and De Bortoli, Valentin and Rainforth, Thomas and Deligiannidis, George and Doucet, Arnaud},
  journal={Proc. Adv. Neural Inf. Process. Syst.},
  volume={35},
  pages={28266--28279},
  year={2022}
}

@article{sun2022score,
  title={Score-based continuous-time discrete diffusion models},
  author={Sun, Haoran and Yu, Lijun and Dai, Bo and Schuurmans, Dale and Dai, Hanjun},
  journal={arXiv preprint arXiv:2211.16750},
  year={2022}
}

@inproceedings{zhang2021causal,
  title={Causal intervention for leveraging popularity bias in recommendation},
  author={Zhang, Yang and Feng, Fuli and He, Xiangnan and Wei, Tianxin and Song, Chonggang and Ling, Guohui and Zhang, Yongdong},
  booktitle={Proc. 44th Int. ACM SIGIR Conf. Res. Develop. Inf. Retrieval},
  pages={11--20},
  year={2021}
}

@article{huang2022exploring,
  title={Exploring consumer online purchase and search behavior: An FCB grid perspective},
  author={Huang, Shiu-Li and Lin, Yi-Hsien},
  journal={Asia Pacific Management Review},
  volume={27},
  number={4},
  pages={245--256},
  year={2022},
  publisher={Elsevier}
}

@incollection{oksendal2003stochastic,
  title={Stochastic differential equations},
  author={{\O}ksendal, Bernt},
  booktitle={Stochastic differential equations: an introduction with applications},
  pages={38--50},
  year={2003},
  publisher={Springer}
}

@inproceedings{lou2024discrete,
  title={Discrete Diffusion Modeling by Estimating the Ratios of the Data Distribution},
  author={Lou, Aaron and Meng, Chenlin and Ermon, Stefano},
  year={2024},
  booktitle={Proc. Int. Conf. Mach. Learn.}
}

@article{meng2022concrete,
  title={Concrete score matching: Generalized score matching for discrete data},
  author={Meng, Chenlin and Choi, Kristy and Song, Jiaming and Ermon, Stefano},
  journal={Proc. Adv. Neural Inf. Process. Syst.},
  volume={35},
  pages={34532--34545},
  year={2022}
}

@inproceedings{song2023consistency,
  title={Consistency Models},
  author={Song, Yang and Dhariwal, Prafulla and Chen, Mark and Sutskever, Ilya},
  booktitle={Proc. Int. Conf. Mach. Learn.},
  pages={32211--32252},
  year={2023},
  organization={PMLR}
}

@inproceedings{geng2024consistency,
  title={Consistency Models Made Easy},
  author={Geng, Zhengyang and Pokle, Ashwini and Luo, Weijian and Lin, Justin and Kolter, J Zico},
  year={2025},
  pages={96638--96666},
  booktitle={Proc. 13th Int. Conf. Learn. Representations}
}

@inproceedings{ ning2023elucidating,
title={Elucidating the Exposure Bias in Diffusion Models},
author={Mang Ning and others},
booktitle={Proc. 12th Int. Conf. Learn. Representations},
pages={21641-21664},
year={2024}
}

@article{zhang2023contrastive,
  title={Contrastive sampling chains in diffusion models},
  author={Zhang, Junyu and Liu, Daochang and Zhang, Shichao and Xu, Chang},
  journal={Proc. Adv. Neural Inf. Process. Syst.},
  volume={36},
  pages={73524--73542},
  year={2023}
}

@inproceedings{yang2023generative,
  title={Generative-contrastive graph learning for recommendation},
  author={Yang, Yonghui and Wu, Zhengwei and Wu, Le and Zhang, Kun and Hong, Richang and Zhang, Zhiqiang and Zhou, Jun and Wang, Meng},
  booktitle={Proc. 46th Int. ACM SIGIR Conf. Res. Develop. Inf. Retrieval},
  pages={1117--1126},
  year={2023}
}

@article{xu2024discrete,
  title={Discrete-state continuous-time diffusion for graph generation},
  author={Xu, Zhe and Qiu, Ruizhong and Chen, Yuzhong and Chen, Huiyuan and Fan, Xiran and Pan, Menghai and Zeng, Zhichen and Das, Mahashweta and Tong, Hanghang},
  journal={Proc. Adv. Neural Inf. Process. Syst.},
  volume={37},
  pages={79704--79740},
  year={2024}
}

@inproceedings{wang-etal-2025-knowledge-graph,
    title = "Knowledge Graph Retrieval-Augmented Generation for {LLM}-based Recommendation",
    author = "Wang, Shijie  and
      Fan, Wenqi  and
      Feng, Yue  and
      Shanru, Lin  and
      Ma, Xinyu  and
      Wang, Shuaiqiang  and
      Yin, Dawei",
    year = "2025",
    booktitle={Proc. Annu. Meet. Assoc. Comput. Linguist.},
    publisher = "Association for Computational Linguistics",
    pages = "27152--27168",
    ISBN = "979-8-89176-251-0",
}

@article{liu2024score,
  title={Score-based generative diffusion models for social recommendations},
  author={Liu, Chengyi and Zhang, Jiahao and Wang, Shijie and Fan, Wenqi and Li, Qing},
  journal={IEEE Trans. Knowl. Data Eng.},
  year={2025},
  volume={37},
  pages={6666-6679},
  publisher={IEEE}
}

@inproceedings{lee2025scone,
  title={SCONE: A Novel Stochastic Sampling to Generate Contrastive Views and Hard Negative Samples for Recommendation},
  author={Lee, Chaejeong and Choi, Jeongwhan and Wi, Hyowon and Cho, Sung-Bae and Park, Noseong},
  booktitle={Proc. 18th ACM Int. Conf. Web Search Data Mining},
  pages={419--428},
  year={2025}
}

@inproceedings{chen2023fairly,
  title={Fairly adaptive negative sampling for recommendations},
  author={Chen, Xiao and Fan, Wenqi and Chen, Jingfan and Liu, Haochen and Liu, Zitao and Zhang, Zhaoxiang and Li, Qing},
  booktitle={Proc. World Wide Web Conf.},
  pages={3723--3733},
  year={2023}
}

@article{qu2024ssd4rec,
  title={Ssd4rec: a structured state space duality model for efficient sequential recommendation},
  author={Qu, Haohao and Zhang, Yifeng and Ning, Liangbo and Fan, Wenqi and Li, Qing},
  journal={arXiv preprint arXiv:2409.01192},
  year={2024}
}

@article{liu2025glprotein,
  title={GLProtein: Global-and-Local Structure Aware Protein Representation Learning},
  author={Liu, Yunqing and Fan, Wenqi and Wei, Xiaoyong and Li, Qing},
  journal={arXiv preprint arXiv:2506.06294},
  year={2025}
}

@inproceedings{fan2023generative,
  title={Generative Diffusion Models on Graphs: Methods and Applications},
  author={Liu, Chengyi and Fan, Wenqi and Liu, Yunqing and Li, Jiatong and Li, Hang and Liu, Hui and Tang, Jiliang and Li, Qing},
  booktitle={Proc. 32nd Int. Joint Conf. Artif. Intell.},
  pages={6702--6711},
  year={2023}
}

@article{he2025graph,
  title={Graph Defense Diffusion Model},
  author={He, Xin and Fan, Wenqi and Wang, Yili and Liu, Chengyi and Miao, Rui and Juan, Xin and Wang, Xin},
  journal={arXiv preprint arXiv:2501.11568},
  year={2025}
}

@article{wang2025graph,
  title={Graph machine learning in the era of large language models (llms)},
  author={Wang, Shijie and Huang, Jiani and Chen, Zhikai and Song, Yu and Tang, Wenzhuo and Mao, Haitao and Fan, Wenqi and Liu, Hui and Liu, Xiaorui and Yin, Dawei and others},
  journal={ACM Trans. Intell. Syst. Technol.},
  volume={16},
  number={5},
  pages={1--40},
  year={2025},
  publisher={ACM New York, NY}
}

@article{qu2025tokenrec,
  title={TokenRec: Learning to Tokenize ID for LLM-Based Generative Recommendations},
  author={Qu, Haohao and Fan, Wenqi and Zhao, Zihuai and Li, Qing},
  journal={IEEE Transactions on Knowledge and Data Engineering},
  year={2025},
  publisher={IEEE}
}

@inproceedings{ning2024cheatagent,
  title={Cheatagent: Attacking llm-empowered recommender systems via llm agent},
  author={Ning, Liang-bo and Wang, Shijie and Fan, Wenqi and Li, Qing and Xu, Xin and Chen, Hao and Huang, Feiran},
  booktitle={Proceedings of the 30th ACM SIGKDD Conference on Knowledge Discovery and Data Mining},
  pages={2284--2295},
  year={2024}
}
